\documentclass[%
 reprint,
superscriptaddress,
nofootinbib,
 amsmath,amssymb,
 aps,
]{revtex4-1}

\usepackage{aas_macros}
\usepackage{graphicx}
\usepackage{dcolumn}
\usepackage{fontawesome}
\usepackage{bm}
\usepackage[normalem]{ulem}

\usepackage{xcolor}

\definecolor{colorLink}{rgb}{0.9,0,0} % red
\definecolor{colorCite}{rgb}{0,0.7,0} % green
\definecolor{colorURL} {rgb}{0,0,0.8} % navy
\usepackage[colorlinks=true,linktocpage=true,linkcolor=colorLink,citecolor=colorCite,urlcolor=colorURL]{hyperref}
\usepackage{dcolumn}% Align table columns on decimal point
\usepackage{physics}
\usepackage{slashed}

\newcommand{\HI}{H\textsc{i}\,\,}
\newcommand{\HII}{H\textsc{i}}
\newcommand{\be}{\begin{equation}}
\newcommand{\ee}{\end{equation}}
\newcommand{\bad}{\begin{aligned}}
\newcommand{\ead}{\end{aligned}}
\newcommand{\lagr}{\mathcal{L}}
\newcommand{\DM}{\mathrm{DM}}

\newcommand{\nH}{n_{\mathrm{H}}}
\newcommand{\vrel}{v_{\mathrm{rel}}}
\newcommand{\mdm}{M_\DM}
\newcommand{\rdm}{R_\DM}

\newcommand{\rwnm}{r_{\mathrm{WNM}}}

\defcitealias{WadFar21}{WF21}

\defcitealias{DiaKap21}{D21} 

\begin{document}

\preprint{APS/123-QED}

\title{Constraining axion and compact dark matter with interstellar medium heating}

\author{Digvijay Wadekar}
\affiliation{School of Natural Sciences, Institute for Advanced Study, 1 Einstein Drive, Princeton, NJ 08540, USA}
%\email{}
\author{Zihui Wang}
\email{zihui.wang@nyu.edu}
\affiliation{
Center for Cosmology and Particle Physics, Department of Physics, New York University, New York, NY 10003
}%

\date{\today}% It is always \today, today,

\begin{abstract}

Cold interstellar gas systems have been used to constrain dark matter (DM) models by the condition that the heating rate from DM must be lower than the astrophysical cooling rate of the gas.
Following the methodology of \citet{WadFar21}, we use the interstellar medium of a gas-rich dwarf galaxy, Leo T, and a Milky Way-environment gas cloud, G33.4-8.0 to constrain DM. Leo T is a particularly strong system as its gas can have the lowest cooling rate among all the objects in the late Universe (owing to the low volume density and metallicity of the gas). Milky Way clouds, in some cases, provide complementary limits as the DM-gas relative velocity in them is much larger than that in Leo T. We derive constraints on the following scenarios in which DM can heat the gas: $(i)$ interaction of axions with hydrogen atoms or free electrons in the gas, $(ii)$ deceleration of relic magnetically charged DM in gas plasma, $(iii)$ dynamical friction from compact DM, $(iv)$ hard sphere scattering of composite DM with gas. Our limits are complementary to DM direct detection searches. Detection of more gas-rich low-mass dwarfs like Leo T from upcoming 21cm and optical surveys can improve our bounds.

% Future measurement of other gas-rich dwarf galaxies...

% Cold interstellar gas systems have been used to constrain dark matter (DM) models by the condition that the heating rate from DM must be lower than the astrophysical cooling rate of the gas. For instance, if DM scatters with baryons or electrons, the recoiling particles can inject heat to the gas and this provides a complementary probe to DM direct detection; compact DM objects traveling through the gas can directly heat the gas via dynamical friction; magnetically charged DM would be decelerated in the gas plasma and the energy loss is converted to heat. In this paper, we exploit a gas-rich dwarf galaxy, Leo T, and a Milky Way-environment gas cloud, G33.4-8.0, to constrain these scenarios. Specifically, we place upper limits on the electron coupling of electrophilic axion DM, the abundance of compact DM objects, and the spatial size of DM states that scatter with baryons geometrically. For magnetically charged black holes, we derive upper bounds on their abundance depending on the mass and charge. Future measurement of other gas-rich dwarf galaxies...
\end{abstract}

\maketitle

%\tableofcontents

\section{Introduction}

There are a number of well-motivated models of dark matter (DM) that feature couplings to Standard Model (SM) particles or self interactions. A popular example is the QCD axion~\cite{Peccei:1977ur,Weinberg78,Wilczek78}, which is natural to have couplings to photons, leptons and nucleons. Such interactions can be potentially detected in laboratory and astrophysical measurements, but are still consistent with cold DM (CDM) at large scale. DM can also be made up of compact objects; these can have macroscopic interactions with ordinary matter. There can also be candidates such as primordial black holes (PBHs)~\cite{Green:2020jor} which emit SM particles by Hawking radiation and can accrete matter around them. Constraining the interactions of DM is critical to both DM model building and instrumental development.

Direct and indirect detection are two particularly important techniques to discover DM interactions. The strategy of direct detection is to look for signals of nucleon (electron) recoil caused by DM-nucleon (-electron) scattering using Earth-based laboratory detectors such as XENON~\cite{XENON:2020rca}. Limits on DM interactions from these experiments, despite being exceedingly stringent, suffer from the overburden effect~\cite{Zaharijas:2004jv}, and do not apply to sufficiently large cross sections. Due to trigger sensitivity, most of the experiments must require DM particles to be heavy enough, typically $> \order{1}$ GeV for DM-nucleon scattering. In addition to laboratory detectors, a variety of astrophysical systems have also been used to probe DM scattering with SM particles and provide complementary limits, such as CMB~\cite{Xu:2018efh}, the population of satellite galaxies~\cite{Nadler:2019zrb}, planets~\cite{Farrar:2005zd} and exoplanets~\cite{Leane:2020wob}. These astrophysical limits are generally weaker than laboratory limits, but have the advantage of evading the overburden effect and also can be applied to much lighter DM particles. In contrast to directly searches, indirect searches look for visible products of DM decay or annihilation. Limits on decay lifetime and annihilation cross section have been derived from X/$\gamma$-ray telescopes~\cite{Essig:2013goa,HESS:2018cbt}, CMB anisotropy~\cite{Slatyer:2016qyl}, CMB spectal distortion~\cite{BolChl20, Ali21}, line-intensity mapping~\cite{Ber21}, dwarf spheroid galaxies~\cite{Fermi-LAT:2016uux} (see however~\cite{Ando:2020yyk}), Lyman-$\alpha$ forests~\cite{Liu:2020wqz} and cosmic rays~\cite{AMS14}.

\begin{figure}[htb]
\centering
\includegraphics[width=0.45\textwidth]{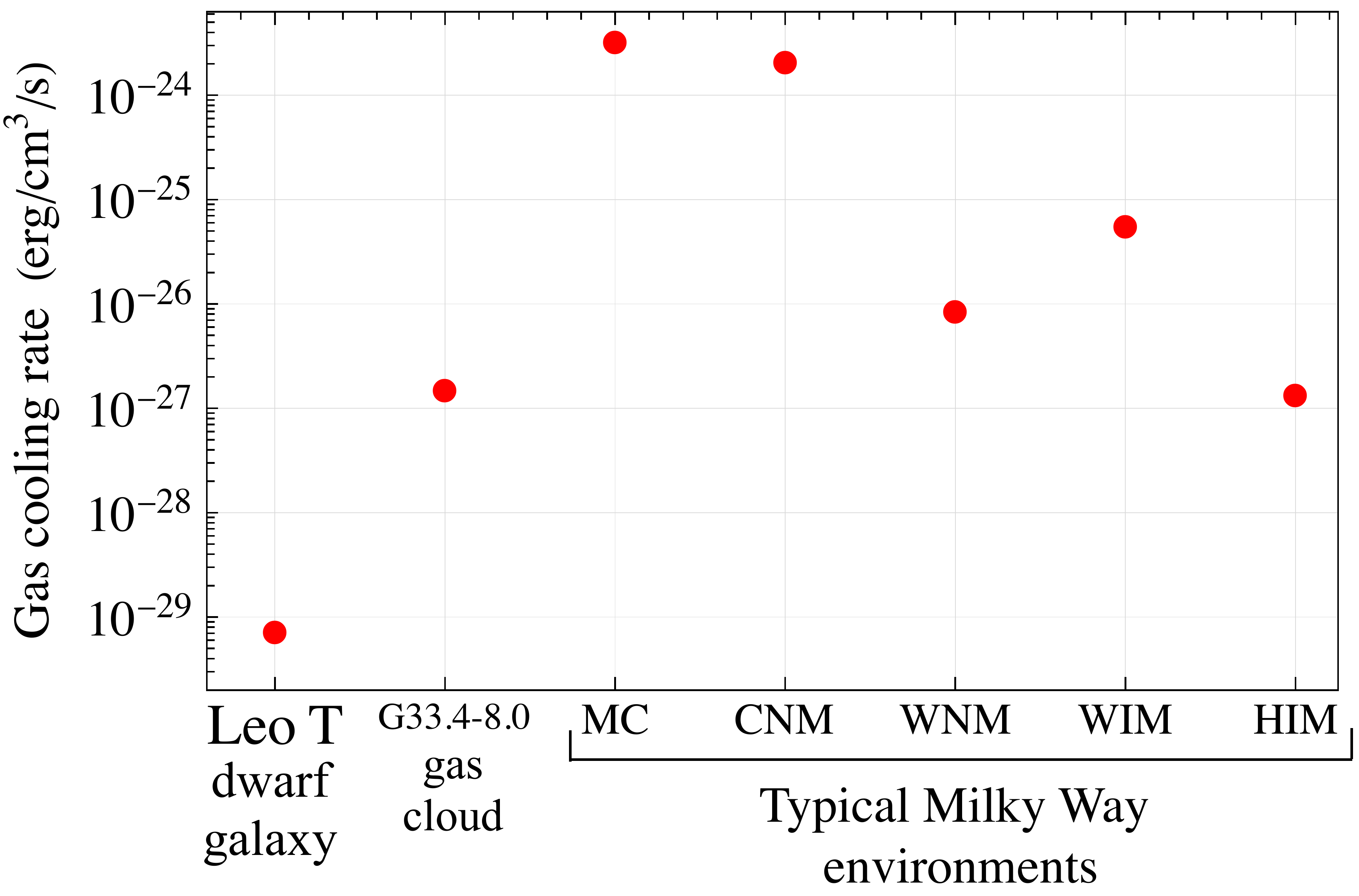}
\caption{Estimates of astrophysical radiative cooling rate of the gas in Leo T, G33.4$-$8.0 cloud, and typical Milky Way interstellar medium phases. Cooling rates generally decrease with lower gas density, metal fraction and temperatures (see section~\ref{sec:property} for further details). Due to its low metallicity and gas volume density, the gas in Leo T has a much lower astrophysical cooling rate than typical systems in the Milky Way (to our best knowledge, gas in Leo T has the lowest cooling rate among astrophysical systems in the late Universe).}
\label{fig:coolrate}
\end{figure}

Recently, it shown that some of the gas-rich astrophysical systems can be used as powerful calorimetric DM detectors %in both direct and indirect scenarios
\cite{Chi90,DubHer15,WadFar21,Bhoonah:2018wmw,BhoBramante19,Bhoonah:2020dzs,Farrar:2019qrv}.
These studies required that DM heat injection rate $\dot{Q}_\mathrm{DM}$ must be lower than the astrophysical cooling rate of the gas $\dot{C}$,
\begin{equation}
    \dot{Q}_\mathrm{DM} \leq \dot{C}\, ,
\label{Qdot}
\end{equation}
otherwise the temperature of the gas would steadily increase (and the ionization state of the gas could also be altered). Systems with low gas cooling rates are therefore more sensitive to energy injections by DM.
To our best knowledge, warm neutral gas in the Leo T dwarf galaxy has the lowest cooling rate among astrophysical systems in the late Universe (see the comparison in Fig.~\ref{fig:coolrate}). This is precisely why Ref.~\cite{WadFar21} used Leo T to constrain heating due to DM.

In particular, Ref.~\cite{WadFar21} derived limits on DM-nucleon/electron scattering cross sections, and the mixing parameter of dark photon DM. Refs.~\cite{LuTak21,LahLu20,Kim20,TakLu21,Tak21b} used Leo T to constrain various heating mechanisms due to primordial black holes (PBH) (e.g., Hawking radiation, accretion disk, outflows and dynamical friction), and obtained upper bounds on the abundance of PBHs. In an earlier work~\cite{Wadekar:2021qae}, we used Leo T to place limits on DM decay and annihilation to $e^+ e^-$ and $\gamma\gamma$ pairs, updating existing limits for $\order{100\, \mathrm{eV}}$ photons and $\order{1\, \mathrm{MeV}}$ electrons.

% A well studied gas-rich dwarf galaxy, Leo T, with an estimated cooling rate $\dot{C} \simeq 7\times 10^{-30}$ $\mathrm{erg}\, \mathrm{cm}^{-3}\, \mathrm{s}^{-1}$~\cite{Kim20,WadFar21}, was used by Ref.~\cite{WadFar21} to constrain heat injection from DM to gas.

%\citet{WadFar21} (hereafter \citetalias{WadFar21}) showed that gas-rich dwarfs can be used to constrain DM models. Because of of their ultra-low gas cooling rates owing to their low metallicity, they are very sensitive to heat injection by a non-standard energy source. \WF\ used a particular well-studied gas-rich dwarf galaxy called Leo~T and required heat exchange between DM and ordinary matter to not exceed the gas cooling rate. This leads to strong limits on
%DM scattering with ordinary matter and on hidden photon dark matter (HPDM). Subsequently, the gas cooling rate of Leo~T has also been used to constrain gas heating due to primordial black holes (PBHs) by Refs.~\cite{LuTak21,LahLu20,Kim20,TakLu21,Tak21b} (the gas heating from PBHs occurs via various mechanisms like radiation from the BH accretion disk, Hawking radiation, BH outflows and dynamical friction).

To set strong bounds on DM, not only should $\dot{C}$ in Eq.~\ref{Qdot} be lower, but $\dot{Q}_\mathrm{DM}$ also should be larger. There are many models of DM where $\dot{Q}_\mathrm{DM}$ increases as a function of DM-gas relative velocity (e.g., in the scenario where cross section of DM-baryon interactions is velocity-independent, $\dot{Q}_\mathrm{DM}\propto v^3_\mathrm{relative}$).
Leo T has low relative velocity between DM and baryons $\sim 17$ km/s, whereas systems in the Milky Way (MW) have $v_\mathrm{relative}\sim 300$ km/s. Therefore, in such scenarios, MW systems can potentially provide stronger limits than Leo T.
A variety of MW gas clouds have therefore been used for constraining DM models such as millicharged DM, asymmetric DM nuggets, DM-nuclei contact interactions, magnetically charged black holes and DM decay/annihilation \cite{Bhoonah:2018wmw,Bhoonah:2020dzs,DiaKap21,Wadekar:2021qae}
% Ref.~\cite{Bhoonah:2018wmw} uses an diffuse neutral gas cloud G1.4-1.8+87 to constrain the heat exchange between millicharged DM and gas (see however~\cite{Farrar:2019qrv}). Refs.~\cite{Bhoonah:2020dzs,DiaKap21,Wadekar:2021qae} have used gas clouds to constrain asymmetric DM nuggets, DM-nuclei contact interactions, magnetically charged black holes and DM decay/annihilation.

% However, it does not guarantee Leo T to give the strongest limits on DM heating. This is because the DM velocity dispersion in MW is $\order{10}$ times larger than in Leo T, and in many scenarios the DM heating rate is velocity dependent. Therefore, in this paper we will use both the Leo T galaxy and a particular MW gas cloud, G33.4-8.0~\cite{Farrar:2019qrv}, to constrain DM heating. Hereafter, we use the phrase, the MW cloud, to refer to G33.4-8.0. The cooling rate of the MW gas cloud is estimated to be $2.1\times 10^{-27}$ $\mathrm{erg}\, \mathrm{cm}^{-3}\, \mathrm{s}^{-1}$~\cite{WadFar21}.

In this paper, we will use both the Leo T galaxy and a robust MW gas cloud, G33.4-8.0~\cite{Farrar:2019qrv}, to constrain DM heating (hereafter, we use the phrase, the MW cloud, to refer to G33.4-8.0). We derive new limits on a few DM models using the interstellar gas heating argument, as well as update certain existing limits that used inaccurate inputs. The paper is organized as follows. In Sec.~\ref{sec:property}, we review the properties of the gas systems. In Sec.~\ref{sec:axion}, we study the heating due to electrophilic axion DM and set limits on the electron coupling. In Sec.~\ref{sec:compact}, we derive limits on heat injection from compact DM objects via dynamical friction, hard sphere scattering and magnetic effects.

%%%%%%%%%%%%%%%%%%%%%%%%%%%%%%%%%%%%%%%%%%%%%%%%%%%%%%%%%%%%%%%%%
\section{Properties of Leo T and Milky Way gas cloud}
\label{sec:property}

In this section, we discuss properties of the astrophysical systems and the formalism for calculating their radiative gas cooling rate $\dot{C}$.
Depending on the temperature and ionization fraction of the gas, interstellar gas systems can be generically classified to five types: molecular clouds (MC), cold neutral medium (CNM), warm neutral medium (WNM), warm ionized medium (WIM), and hot ionized medium (HIM) \cite{Dra11}. 
The gas in the inner part of the Leo T galaxy is dominated by WNM with $T \simeq 6100$ K~\cite{leoObsOld08, adams17}.
The spatial profile of DM, hydrogen, and free electrons in Leo T was determined by Ref.~\cite{faerman13} upon fitting a hydrostatic model to \HI column density observations and assuming that the DM follows a Burkert (cored) profile \cite{Bur95}:
\begin{equation}
\rho_\mathrm{DM} = \frac{\rho_0}{(1+ \frac{r}{r_s})(1+ \frac{r^2}{r^2_s})}
\label{eq:burkert}
\end{equation}
where $r$ is the radial distance from the halo center, $r_s$ is the scale radius and $\rho_0$ is the central core density. Recent observations suggest the presence of roughly constant density cores in most of the low-mass dwarf galaxies, therefore the choice of Burkert profile for Leo T is well-motivated. Furthermore, assuming a cuspy profile (e.g., NFW) will give tighter constraints on DM, hence our assumption of cored profiles is conservative. The best-fit profiles from Ref.~\cite{faerman13} are shown in Fig.~\ref{fig:LeoTobs} (corresponding to $\rho_0 = 3.9$ GeV/cm$^3$ and $r_s=0.7$ kpc for the DM halo), and we adopt them for our calculations.\footnote{The $2\sigma$ errors on Leo T halo parameters reported in Ref.~\cite{faerman13} leads to a $\lesssim 10\%$ variation to the DM heating rate~\cite{Wadekar:2021qae} and hence only weakly impact the results of this paper.}

A widely-used approximate formula to calculate the cooling rate is~\cite{grackle}
\begin{equation}
\dot{C}=n^2_\textup{H}\, \Lambda (T) 10^\textup{[Fe/H]}\, ,
\label{eq:grackle}
\end{equation}
where $\nH$ is the number density of hydrogen in the gas, $T$ is the temperature, $\textup{[Fe/H]}$ is the metallicity relative to the Sun, and $\Lambda(T)$ is a monotonically increasing function of $T$ taken from~\cite{grackle} (also known as the `cooling function', see Fig.~7 of \cite{WadFar21}).

For WNM of Leo T, Eq.~\ref{eq:grackle} gives $\dot{C}\sim 4 \times 10^{-30}$ $\mathrm{erg}\, \mathrm{cm}^{-3}\, \mathrm{s}^{-1}$.
However, a more accurate calculation of $\dot{C}$ for Leo T was performed in Ref.~\cite{Kim20}; we conservatively use their result throughout this paper: $\dot{C} = 7\times 10^{-30}$ $\mathrm{erg}\, \mathrm{cm}^{-3}\, \mathrm{s}^{-1}$ (see Appendix~A of \cite{Wadekar:2021qae} for a detailed discussion of the differences between the two approaches). Note that using the $2\sigma$ conservative value of the WNM temperature of Leo T would increase $\dot{C}$ only by a factor of two ($\dot{C} \simeq 14.6\times 10^{-30}$ $\mathrm{erg}\, \mathrm{cm}^{-3}\, \mathrm{s}^{-1}$ corresponding to $T \simeq 7552$ K~\cite{Wadekar:2021qae}), and therefore impacts our DM constraints weakly.

For the MW cloud, we follow Ref.~\cite{WadFar21} and take the DM density to be 0.64 GeV/cm${}^{3}$, \HI density $0.4$/cm${}^{3}$, and its cylindrical coordinates relative to the center of the Milky Way being $R= 4.68\pm 0.41$ kpc and $z = 1\pm 0.28$ kpc. The cooling rate of WNM of the MW gas cloud is estimated to be $2.1\times 10^{-27}$ $\mathrm{erg}\, \mathrm{cm}^{-3}\, \mathrm{s}^{-1}$~\cite{WadFar21} using Eq.~\ref{eq:grackle}. In Fig.~\ref{fig:coolrate}, we show a comparison of the cooling rate of Leo T and the MW cloud and also include a rough estimate of typical cooling rates of different ISM phases of the Milky Way. We leave further discussion on the Milky Way ISM phases to Appendix~\ref{app:cooling}. We see that cooling rates of systems in the Milky Way are generally a few orders of magnitude larger than the that of Leo T.

%One important point to note is that the cooling function increases monotonically with temperature.

An important quantity for calculating the heating rate due to DM is the velocity of DM relative to the gas. In Leo T, the gas has no observable rotation. The velocity dispersion of both the gas and DM is observationally determined\footnote{the velocity dispersion of DM particles is assumed to be roughly similar to that of stars (which is observed to be $7.6^{+2.3}_{-1.7}$ km/s \cite{Zou21,SimonGeha07}), as both are nearly collisionless and trace the underlying potential.} to be $\sigma_v \sim 7$ km/s~\cite{leoObsOld08, adams17}. We thus assume that velocities of gas and DM particles in Leo T approximately follow identical Maxwell distributions and write the distribution of DM-gas relative velocity $\vrel$ as~\cite{TakLu21}
\begin{equation}
    f(\vrel) = \frac{1}{N_{\mathrm{esc}}}\int_0^{v_{\mathrm{esc}}} dv \, \frac{v \vrel}{\pi \sigma_v^4} e^{-\frac{v^2}{2\sigma_v^2}}(e^{-\frac{(\vrel-v)^2}{2\sigma_v^2}}-e^{-\frac{(\vrel+v)^2}{2\sigma_v^2}}),
\label{eq:vrelLeoT}
\end{equation}
We conservatively assumed that the escape velocity $v_{\mathrm{esc}}$ is 23.8 km/s 
%\DW{would be good to show derivation here (I guess this is estimated assuming a dynamical mass $M_\mathrm{dyn}\sim 10^7 M_\odot$)} 
and the normalization constant $N_{\mathrm{esc}}$ is set by the condition $\int_{0}^{2 v_{\mathrm{esc}}} d\vrel\, f(\vrel) = 1$. 
The estimated escape velocity follows from $v_{\mathrm{esc}} = \sqrt{2GM/\rwnm}$, where we take the halo mass: $M = 2.3\times 10^7 M_\odot$ by integrating the DM profile to $\rwnm=0.35$ kpc. Note that we have made a conservative estimate for the escape velocity as the DM halo of Leo T is expected to continue far beyond $R = 0.35$ kpc and one would typically expect dwarf spheroidals like Leo T to have $M_\mathrm{dyn}\sim 10^9 M_\odot$. A higher escape velocity would shift the center of the velocity distribution to the higher end, and therefore leads to stronger constraints if the DM heating rate increases with velocity. 
A more realistic estimate for the Leo T escape velocity can be derived from $v_{\mathrm{esc}}(r) = \sqrt{2|\Phi(r) - \Phi(3\,r_{340})|}$~\cite{Piffl:2013mla}, where $\Phi$ is the gravitational potential and $r_{340}$ is the radius where the enclosing density is 340 times the critical density ($3\, r_{340}$ is assumed to be the boundary of the halo). Using the halo model of Leo T given in Eq.~(\ref{eq:burkert}), we find $r_{340} \sim 18$ kpc and $v_{\mathrm{esc}} \sim 62$ km/s at $r=0.35$ kpc. In later sections, we use both 23.8 km/s and 62 km/s as the escape velocity to derive DM limits, but find the dependence of the limits on $v_\mathrm{esc}$ to be negligible.
%Evaluating $\expval{\vrel} \equiv \int d\vrel \vrel f(\vrel)$ yields $\expval{\vrel} = 15.3$ km/s. Note that here the upper bound of integration is $\vrel = 2v_{\mathrm{esc}}$.
%Evaluating $\expval{\vrel} \equiv \int d\vrel \vrel f(\vrel)$ yields $\expval{\vrel} = 15.8$ km/s, coinciding with $\sqrt{2}$ times the average of individual Maxwell distribution. 
%By comparison, the speed of sound in the \HI gas is $\sim$ 9 km/s, and the DM objects are therefore mildly supersonic.

The MW cloud rotates about the galactic center at a bulk velocity $v_b = 220$ km/s and the DM velocity dispersion is $\sigma_v = 124.4$ km/s~\cite{Pidopryhora2015}. The escape velocity is approximately $v_{\mathrm{esc}} \sim 600$ km/s~\cite{Piffl:2013mla}. Then, $f(\vrel)$ can be calculated by~\cite{TakLu21}
\begin{equation}
     f(\vrel) = \frac{1}{N_{\mathrm{esc}}} \frac{\vrel}{\sqrt{2\pi} \sigma_v v_b} (e^{-\frac{(\vrel-v_b)^2}{2\sigma_v^2}}-e^{-\frac{(\vrel+v_b)^2}{2\sigma_v^2}}).
\label{eq:vrelMW}
\end{equation}
Again, we fix the normalization constant by requiring $\int_0^{v_{\mathrm{esc}}+v_b} d\vrel\, f(\vrel) = 1$.

%%%
\section{Limits on Axion DM}
\label{sec:axion}

In an earlier work~\cite{Wadekar:2021qae}, we set limits on the photon coupling of DM axions based on the gas temperature in Leo T. In the current paper, we derive limits on the coupling of axions to electrons. We consider the scenario of electrophilic axions, where axions only couple to electrons (not photons) at the tree level through the following Lagrangian 
\begin{equation}
\lagr = -\frac{1}{2}m_a a^2 - g_{ae} a \bar{\psi}_{e} \gamma^5 \psi_{e}.
\end{equation}
Electrophilic axions can heat the gas in a number of ways.
\begin{itemize}
    \item Analogous to photoelectric effect, axions can be absorbed by atoms and generate electron recoil via axioelectric effect~\cite{dimopoulos86,pospelov08}. Subsequently, the recoiling electrons can deposit their kinetic energy to the gas. For Leo T, we restrict our discussion to hydrogen atoms only because they are the major component of the WNM. As the axion is totally absorbed by hydrogen, the kinetic energy of the recoiling electron is equal to the axion mass minus the electron binding energy. Thus, the volume averaged heat injection rate can be modeled by
\begin{equation}
    \dot{Q} = \frac{\sigma_{ae} \vrel E_{\mathrm{heat}}}{m_a \rwnm^3/3} \int dr\, \rho_a(r) \nH(r) ,
\label{eq:axionheat}
\end{equation}
where the integral is performed on the spatial region of the WNM from $r=0$ to $\rwnm=0.35$ kpc, $\nH$ is the number density of neutral hydrogen, $\sigma_{ae}$ is the axioelectric cross section, and $E_{\mathrm{heat}} = m_a f_e(m_a)$ is the energy deposited by the recoiling electron. Here the function $f_e$ gives the heating efficiency of electrons with kinetic energy $m_a$ and can be found by Eq.~(16) in Ref.~\cite{Kim20}. We leave further elaboration on the calculation of the heating rate in Appendix~\ref{app:axion}.

    \item The coupling of axions to electrons allows the decay of axions to two photons via a triangle loop of electrons. For $m_a < m_e$, the one-loop effective coupling $g_{a\gamma\gamma}$ (which sets the decay lifetime) is given by~\cite{pospelov08,Ferreira:2022egk}
    \begin{equation}
        g_{a\gamma\gamma} = \frac{\alpha g_{ae}}{\pi m_e} (1 - x^{-2} \arcsin^2 x),
    \end{equation}
    where $x \equiv m_a/(2m_e)$. We then use the methodology given in section~3B of Ref.~\cite{Wadekar:2021qae} to calculate the heating of gas in Leo T as a function of photon energy.
    
    \item The WNM in Leo T also contains a small amount of free electrons (see Fig.~\ref{fig:LeoTobs}). Axions can interact with these free electrons via inverse Compton scattering $ae\to e\gamma$. However, as the number density of free electrons in \HI gas of Leo T is small (the ionization fraction is at the percent level, see Fig.~1 of \cite{WadFar21}), we will thus neglect the heat injection due to inverse Compton scattering. Eventually, this gives a conservative estimate of the total heating rate from axion DM.
\end{itemize}

\begin{figure}[htb]
    \centering
    \includegraphics[width=0.45\textwidth]{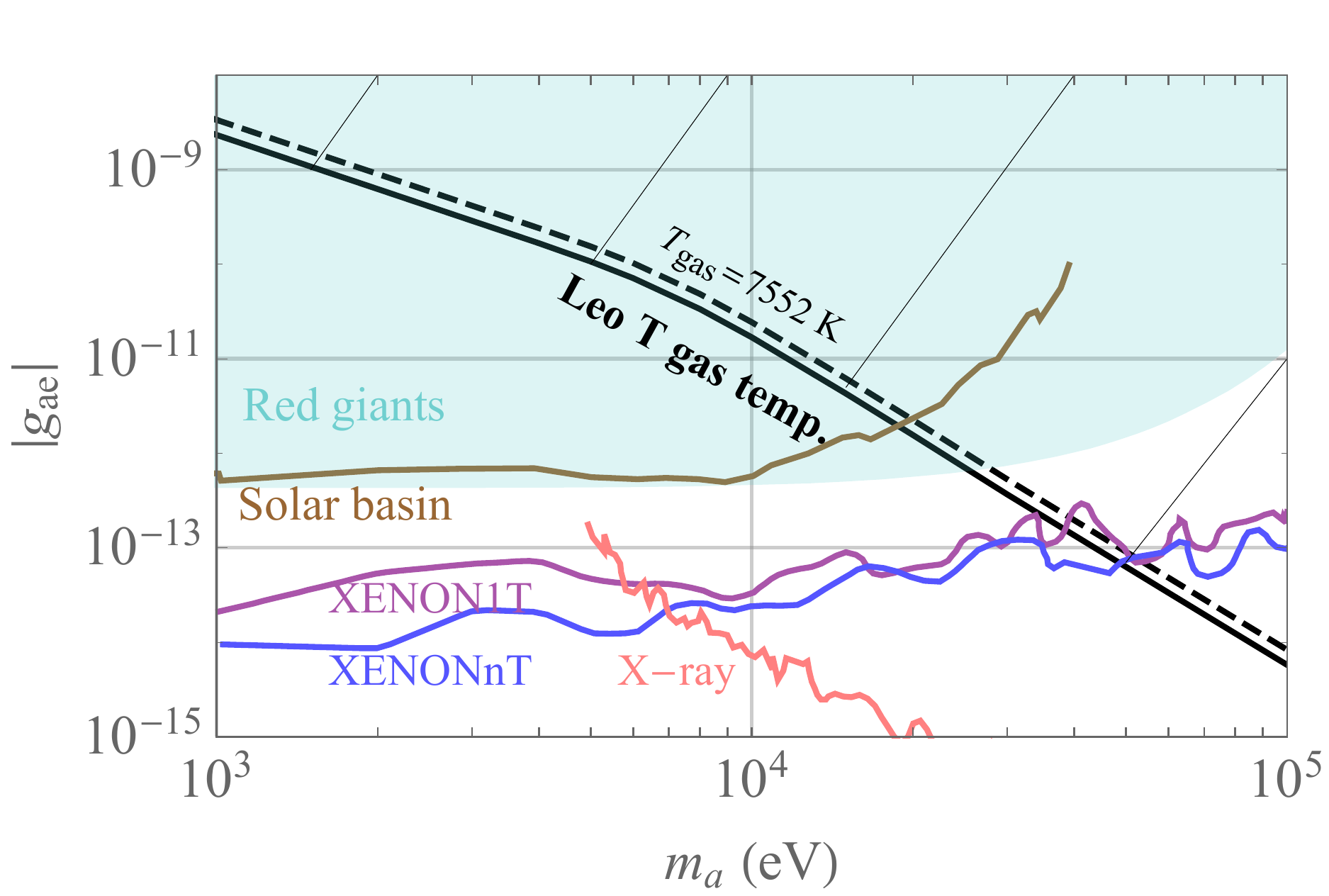}
    \caption{Upper limits on the electron coupling of DM axions. The solid black line is derived from the gas temperature of Leo~T by requiring heating rate from DM to be less than the astrophysical cooling rate of the gas. We also show the effect
of changing the observational estimate of the gas temperature from $T$=6100 K to the 2$\sigma$ conservative value: $T$=7552 K. Other limits shown in the plot include red giants (cyan)~\cite{Capozzi:2020cbu}, XENON1T (violet)~\cite{XENON:2020rca}, XENONnT (blue)~\cite{XENONnT}, solar basin (brown)~\cite{VanTilburg:2020jvl,solarbasin2} and X-ray (red)~\cite{Ferreira:2022egk, Langhoff:2022bij}. }
    \label{fig:gae}
\end{figure}

Requiring the total heating rate to be lower than the cooling rate produces an upper bound on $g_{ae}$. We show the result for $1\, \mathrm{keV} \leq m_a \leq 100 \, \mathrm{keV}$ by the black curve in Fig.~\ref{fig:gae}. At $m_a=100$~eV, the limit weakens to $g_{ae} \lesssim 10^{-6}$. The $2\sigma$ conservative temperature of WNM in Leo T: 7552 K~\cite{adams17} leads to a slightly larger cooling rate $\dot{C} = 14.6\times 10^{-30}$ $\mathrm{erg}\, \mathrm{cm}^{-3}\, \mathrm{s}^{-1}$ \cite{Wadekar:2021qae}, and the corresponding upper limit on $g_{ae}$ is given by the dashed line. We also show limits from red giants (cyan)~\cite{Capozzi:2020cbu}, XENON1T (violet)~\cite{XENON:2020rca}, XENONnT (blue)~\cite{XENONnT}, solar basin (brown)~\cite{VanTilburg:2020jvl,solarbasin2} and X-ray (red)~\cite{Ferreira:2022egk, Langhoff:2022bij}. Other similar limits are not included in the plot and we refer readers to Ref.~\cite{AxionLimits, Adams:2022pbo} for a complete compilation of existing limits on $g_{ae}$.\footnote{\url{https://github.com/cajohare/AxionLimits}} 

Importantly, we stress that limits from most Earth-based detection experiments (e.g., XENON1T, and also the solar basin limit which is recast from XENON1T) are subject to overburden effects~\cite{Zaharijas:2004jv} (i.e., if the coupling is too strong, the DM particles will be scattered by the earth's crust or the atmosphere before they can reach the detectors). Therefore the direct detection limits may not apply to sufficiently large $g_{ae}$. Astrophysical limits naturally evades this limitation and are thus a valuable complement to laboratory limits, excluding the parameter space of large $g_{ae}$. Finally, we remark that the Leo T limit, as well as XENON1T and X-ray limits, require the axions to be DM,\footnote{The relic abundance of keV-scale axions may be achieved by misalignment with a dark confining gauge group~\cite{Foster:2022ajl} or the decay of inflaton~\cite{Lee:2014xua}.} whereas stellar cooling and solar basin limits do not rely on the DM assumption.

%%%
\section{Limits on compact DM}
\label{sec:compact}
In this section, we constrain various models of compact DM. We first present an overview of models and then study the heating mechanism due to dynamical friction, hard sphere scattering and magnetic charges. 

%%%
\subsection{Models of compact DM}
The landscape of feasible DM masses ranges from $10^{-22}$ eV ultralight bosons to compact objects with mass scales comparable to the Sun. In astrophysics, an important quantity associated with compact objects is the compactness, defined as the ratio of mass to radius. Below we describe four generic classes of compact DM objects in order of decreasing compactness, \emph{primordial black holes, composite DM, exotic compact objects} and \emph{subhalos}.

$(i)$ Among all models of compact DM objects, primordial black holes (PBHs) are the most widely studied. PBHs can be created by primordial density fluctuations~\cite{zeldovich67,hawking71, carr75}, and if heavy enough ($>10^{15}$~g), they can survive from Hawking evaporation to the present day and behave like DM~\cite{Chapline:1975ojl}. For recent reviews, see e.g.,~\cite{Carr:2016drx, Green:2020jor}. 
%The overdensity required for PBH formation can be realized by, for example, inflationary perturbations~\cite{Ivanov:1994pa, Garcia-Bellido:1996mdl}, cosmic strings~\cite{Polnarev:1988dh}, domain walls~\cite{ Rubin:2000dq}, and the collapse of solitonic states~\cite{Cotner:2016cvr}. Besides, PBHs with electric and magnetic charges have been recently investigated and these scenarios could lead to stable Planck-scale black hole relics~\cite{Maldacena:2020skw,Lehmann:2019zgt,Bai:2019zcd}.
%PBHs can also carry electric and magnetic charges~\cite{Maldacena:2020skw}. Recently, it is argued that extremity can protect small charged black holes from decay, allowing Planck-scale black hole relics to be stable and contribute to DM~\cite{Lehmann:2019zgt,Bai:2019zcd}.

$(ii)$ Composite state of dark sector particles could arise from dark sector interactions, leading to the formation of dark atoms or dark nuclei (see e.g.~\cite{Wise:2014jva,Chacko:2005pe,Kaplan:2009de,Kaplan:2011yj,Chacko:2018vss,Cline:2021itd, Krnjaic:2014xza,Detmold:2014qqa, Foot:2014uba, Foot:2016wvj}). It is also shown that first-order phase transitions in the early Universe can produce composite DM objects such as quark nuggets~\cite{Witten:1984rs,Bai:2018dxf,Hong:2020est}. Last but not least, composite DM objects could appear as solitons like Q-balls~\cite{Coleman:1985ki,Kusenko:1997si}.
%A variety of DM models feature ultraheavy composite and macroscopic states, collectively known as composite DM. Broadly speaking, composite DM is the composite structure of dark sector particles formed from dark sector interactions, cosmological phase transitions or topological and non-topological solitons. For instance, a simple dark sector containing fermions $\chi$ and bosonic mediators $\phi$ with Yukawa interactions can lead to the formation of large stable bound states of $\chi$~\cite{Wise:2014jva}. In presence of electromagnetic- or QCD-like gauge interactions in the dark sector as motivated by Mirror Twin Higgs models~\cite{Chacko:2005pe}, dark atoms or atomic DM can play the role of collisionless cold DM at large scales and self-interacting DM at small scales~\cite{Kaplan:2009de,Kaplan:2011yj,Chacko:2018vss,Cline:2021itd}. It is also possible that the early Universe could undergo a phase of dark nucleosynthesis to produce dark nuclei~\cite{Krnjaic:2014xza,Detmold:2014qqa}, generating more complex structures in the dark sector. Models of composite DM are also extensively discussed in other scenarios, such as quark nuggets~\cite{Witten:1984rs,Bai:2018dxf} and Fermi balls~\cite{Hong:2020est} from first-order confinement phase transitions, and Q-balls~\cite{Coleman:1985ki,Kusenko:1997si} from the cosmological evolution of supersymmetric particles.

$(iii)$ Exotic compact objects (ECOs) are gravitationally-bound bodies of dark sector particles, stabilized by quantum pressure or self repulsion. The size of an ECO can vary between an asteroid and a star. Boson stars~\cite{kaup68,colpi86, Eby:2015hsq,Croon:2018ybs}, and in particular axion stars~\cite{Kolb:1993zz,Visinelli:2017ooc}, are well known examples of ECOs. The similar idea has been recently extended to vector bosons~\cite{Gorghetto:2022sue}. Another possibility for ECO formation is through the complexity in the dark sector. If the dark sector has dissipative interactions similar to SM, there can be viable mechanisms to form mirror stars~\cite{Mohapatra:1996yy,Foot:2000vy,Berezhiani:2003xm,Curtin:2019ngc,Hippert:2021fch} and other ECOs~\cite{Chang:2018bgx,Dvali:2019ewm}.

$(iv)$ DM subhalos are halo-like objects that are spatially more diffuse than ECOs. Many DM models predict the existence of DM subhalos. For example, the smallest possible DM halo is shown to be $10^{-12}$~$M_\odot$~\cite{Profumo:2006bv} in WIMP DM models. Models of QCD axions and axion-like particles also predict the formation of miniclusters and minihalos~\cite{Kolb:1993zz,Fairbairn:2017sil,Buschmann:2019icd, Arvanitaki:2019rax, Xiao:2021nkb}. In early matter domination cosmology, asteroid-mass DM microhalos could be created~\cite{Nelson:2018via, Erickcek:2020wzd, Blinov:2021axd}. A final example is ultracompact minihalos surrounding PBHs~\cite{Ricotti:2007au,Eroshenko:2016yve,Nakama:2019htb,Hertzberg:2020kpm}.

Observational signatures of all these types of compact DM objects can be generally attributed to gravitational and non-gravitational interactions. Gravitational probes include lensing~\cite{Kolb:1995bu,MACHO:1998qtf, Niikura:2017zjd, Niikura:2017zjd, witt94, Fairbairn:2017dmf, Bai:2018bej, Montero-Camacho:2019jte, Smyth:2019whb, Croon:2020wpr, Croon:2020ouk, Bai:2020jfm, Fujikura:2021omw, Dai:2019lud, barnacka12, Katz:2018zrn, Ricotti:2009bs, li2012, VanTilburg:2018ykj, Mondino:2020rkn}, pulsar timing~\cite{Dror:2019twh,Ramani:2020hdo}, accretion~\cite{Ali-Haimoud:2016mbv, Bai:2020jfm, Serpico:2020ehh}, dynamical friction~\cite{Carr_1999, Brandt:2016aco,Koushiappas:2017chw, Lu:2020bmd, TakLu21}, and gravitational waves~\cite{Bird:2016dcv,Giudice:2016zpa,Grabowska:2018lnd,Croon:2018ybs,Hertzberg:2020dbk,Diamond:2021dth,Marfatia:2021twj, Croon:2022tmr}. These probes primarily depend on the mass of compact objects and thus are typically model independent (see however~\cite{Croon:2020ouk,Croon:2020wpr,Bai:2020jfm,Fujikura:2021omw,Dror:2019twh,Ramani:2020hdo, Bai:2020jfm, Croon:2022tmr} where dependence on the spatial size of compact objects are investigated). In Sec.~\ref{sec:df}, we will evaluate the sensitivity of dynamical friction constraints to the size of compact objects. On the other hand, depending on the model, compact DM objects could have non-gravitational signatures. For example, they could scatter with SM particles that saturates the geometric cross section~\cite{Witten:1984rs,DeRujula:1984axn,Bhoonah:2020dzs,Blanco:2021yiy}. In some models, compact DM objects carry charges or couple to photons, allowing them to produce electromagnetic signals~\cite{Ge:2017idw,Lehmann:2019zgt,Bai:2019zcd,Maldacena:2020skw,Kritos:2021nsf,Hertzberg:2020dbk,Amin:2020vja,Eby:2021ece}.
%Gravitational probes, including lensing~\cite{Kolb:1995bu,MACHO:1998qtf, Niikura:2017zjd, Niikura:2017zjd, witt94, Fairbairn:2017dmf, Bai:2018bej, Montero-Camacho:2019jte, Smyth:2019whb, Croon:2020wpr, Croon:2020ouk, Bai:2020jfm, Fujikura:2021omw, Dai:2019lud, barnacka12, Katz:2018zrn, Ricotti:2009bs, li2012, VanTilburg:2018ykj, Mondino:2020rkn}, pulsar timing~\cite{Dror:2019twh,Ramani:2020hdo}, accretion~\cite{Ali-Haimoud:2016mbv, Bai:2020jfm, Serpico:2020ehh}, and dynamical friction~\cite{Carr_1999, Brandt:2016aco,Koushiappas:2017chw, Lu:2020bmd, Takhistov:2021aqx}, are primarily dependent on the mass of compact objects and thus are robust methods. Recently it is shown that microlensing, pulsar timing and accretion constraints have non-trivial dependence on the spatial size of compact objects~\cite{Croon:2020ouk,Croon:2020wpr,Bai:2020jfm,Fujikura:2021omw,Dror:2019twh,Ramani:2020hdo, Bai:2020jfm}. In Sec.~\ref{sec:df}, we will evaluate the sensitivity of dynamical friction constraints to the size of compact objects. Gravitational waves from binary mergers of compact DM objects are another type of gravitational probes~\cite{Bird:2016dcv,Giudice:2016zpa,Grabowska:2018lnd,Croon:2018ybs,Hertzberg:2020dbk,Diamond:2021dth}, but in general the merger rate and the spectrum of gravitational waves are very much model dependent.

%----------------------------------------------------------------------------------------------------------------------------------
\subsection{Magnetic primordial black holes}
\label{sec:magnetic_PBH}

The observed properties of gas in Leo~T have been used to constraining in PBHs \cite{LuTak21,Kim20,LahLu20,TakLu21}. BHs passing through \HI gas can transfer heat to the gas due to various mechanisms like Hawking radiation, dynamical friction, radiation from gas accretion and BH outflows. One can therefore use the cooling rate of gas in Leo T to place limits on heating due to primordial BHs.

In this paper, we focus on magnetically charged BHs.
Uncharged BHs below the mass $\sim 10^{15}$g have a lifetime smaller than the age of the Universe (based on their decay via Hawking radiation).
Charged BHs, however, stop decaying when they get closer to extremality. Therefore, even very small extremal PBHs can survive until the present.

Primordial BHs with magnetic charges (MBH) can be produced by PBHs absorbing magnetic monopoles in the early universe. Poisson fluctuations in the number density of magnetic monopoles can lead to the PBHs acquiring a net magnetic charge. Note electric charge on a BH can be neutralized by accretion of $e^+/e^-$ from pairs which are produced in the electric field outside the BH, but magnetic charge cannot be neutralized by accretion of standard model particles. It is worth noting that there are other ways of creating stable charged BHs, in which they are charged under a dark $U$(1) gauge symmetry and the corresponding dark fermion is much heavier than the electron \cite{Bai:2019zcd}. Extremal magnetic BHs have been shown to have interesting phenomenological effects \cite{BaiBer20, DiaKap21, Mal20}. The spectrum of quasi-normal modes of such charged BHs has been thoroughly investigated and could be probed with future gravitational wave detectors \cite{Zim16,Ber09}.

We will utilize the fact that compact magnetic objects traveling through astrophysical plasmas get decelerated and transfer heat to the plasmas as a result \cite{Mey85,HamSar83,DiaKap21}.
\citet{DiaKap21} (hereafter \citetalias{DiaKap21}) recently derived upper bounds on the possible fraction of DM composed of MBHs. They required that the energy deposited by primordial BHs passing through Milky Way \HI clouds to not exceed their cooling rate. Here, we use the cooling rate of WNM of Leo T to derive a similar constraint on MBHs.

We write the charge of a magnetic BH of mass $M$, in natural units, as \cite{BaiBer20, Mal20}
\be
Q\equiv q\, Q_\mathrm{extremal} = q\, \sqrt{4 \alpha} \frac{M}{M_\textup{pl}}
\label{eq:q}
\ee
where $M_\textup{pl}$ (=1.22$\times 10^{19}$ GeV)  is the Planck mass and $q$ is a dimensionless ratio of the BH charge compared to the extremality case.

%We will adopt the formalism of Ref.~\cite{Mey85} in this section and show the relevant steps of the calculation.
In Leo T, the velocity of the dipole $v$ is less than the electron thermal velocity of the plasma ($\sqrt{2\, T/m_e}$). The heat transferred by an individual object is then given by \cite{Mey85}
\be\begin{split}
\frac{dE}{dt}  =& -\frac{4 \pi^{1/2} n_e}{3\sqrt{2 T m_e}} \left[ \ln (4\pi n_e \lambda^2_D l) +\frac{2}{3} \right] Q^2 v^2\\
=& -10^{-17}\left(\frac{n_e}{\mathrm{cm}^{-3}}\right) \left(\frac{f\, \rho_{\DM}}{\mathrm{GeV}/\mathrm{cm}^{3}}\right) \left(\frac{v}{\mathrm{km/s}}\right)^2\\
&\times \left(\frac{q}{1}\right)^2 \left(\frac{M}{10^{10}\mathrm{g}}\right) \left[ \ln (4\pi n_e \lambda^2_D l) +\frac{2}{3} \right]
\end{split}\ee

where $n_e$ is the electron density, $f$ is the fractional relic density of EMBHs, $\lambda_D$ is the Debye length is given by

\be
\lambda_D = \sqrt{\frac{T}{4\alpha_\textup{em}\pi (\sum_i Z^2_i n_i)}}\, ,
\label{eq:Debye}
\ee
and the attenuation length in the plasma is given by
\begin{equation}
l = \left( \frac{2T_e}{\pi m_e} \right)^{1/4} \frac{1}{v^{1/2}w_p}
\end{equation}
where $w_p = \sqrt{\frac{4\pi n_e \alpha}{m_e}}$ is the plasma frequency.

\begin{figure}
\centering
\includegraphics[scale=0.5,keepaspectratio=true]{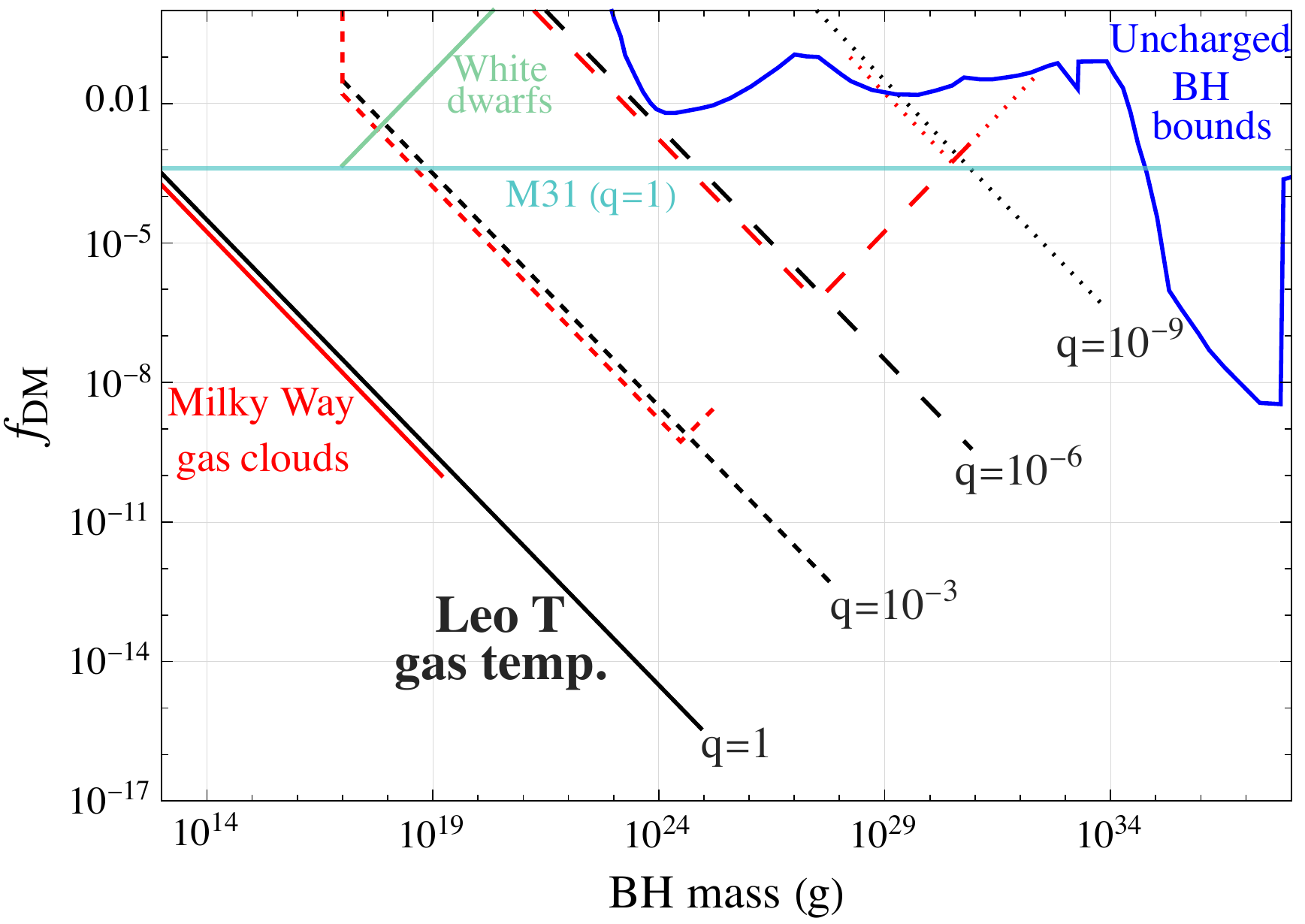}
\caption{Constraints on fraction of black holes with magnetic charge $q$. charged black holes from the Leo~T dwarf galaxy (black), heating of Milky Way's interstellar medium (ISM) (red) \cite{DiaKap21}. Parker bound from the Andromeda galaxy \cite{BaiBer20}. We also show already existing bounds from Ref.~\cite{Car21} for uncharged BH that apply to MBHs.}
\label{fig:EMBH}
\end{figure}

We show the bounds from Leo T in Fig.~\ref{fig:EMBH} alongside the constraints from Milky Way (MW) clouds by \citetalias{DiaKap21}.
%The constraints from \citetalias{DiaKap21} taper off at the point where
The kinks in the red lines correspond to the case when BHs in the MW halo do not pass through the clouds enough. One advantage that Leo T has over the MW clouds is \HI is much more widely distributed in it ($r_\mathrm{WNM}\sim 0.35$ kpc for Leo T, whereas the clouds used in \citetalias{DiaKap21} have sizes $\mathcal{O}$(pc)). Note that the Leo T bound also ultimately cuts off at the point when less than 1 MBH can exist within $r_\mathrm{WNM}$, i.e. $f_\textup{DM} M_\mathrm{halo}<M_\mathrm{BH}$ ($M_\mathrm{halo}\equiv \int_0^{r_\mathrm{WNM}} d^3r \rho_\mathrm{DM}$ is the DM mass enclosed within $r_\mathrm{WNM}$).
%Thus the extent of constraints is improved upon using Leo T.
We have also checked that the energy lost by MBHs in Leo T is too small to affect their orbits within the age of the Universe.

% There could be additional constraints in the region in Fig.~ in case EMBH can catalyze nuclear decay (see)
% There are some additional model-dependent  constraints in Fig.~\ref{fig:EMBH}.
% If magnetic BHs fall inside white dwarfs, they can get trapped due to frictional forces and ultimately consume the white dwarf.  used this fact to set limits on the MBH abundance based on the observed abundance of white dwarfs.
%The bounds from the white dwarfs are dependent on whether EMBHs catalyze proton decay (see Ref.~\cite{DiaKap21} for further discussion).

%%%
\subsection{Dynamical friction} \label{sec:df}
Dynamical friction (DF) is the effect of the net gravitational interactions from a cloud of lighter bodies on a massive object that is traversing the cloud~\cite{chand43}. As a result, the massive traversing object is slowed down and the light bodies in the cloud are accelerated by the gravitational pull. Previously, DF has been used to constrain PBHs traveling in astrophysical environments~\cite{Carr_1999,Brandt:2016aco, Koushiappas:2017chw, Lu:2020bmd, TakLu21}. The argument is that DF would cause stars in star clusters (or gas particles in interstellar gas) to gain energy and increase their velocity dispersion (or temperature) beyond the observed values. 
%hence destabilize the equilibrium of the system.
In this section, we generalize the methodology of Refs.~\cite{Lu:2020bmd, TakLu21} to study DF constraints on spatially extended compact DM objects using gas temperature of Leo T.

The energy loss rate of a compact DM object with mass $\mdm$ and size $\rdm$ due to DF in a gaseous medium is given by~\cite{Binney08, Ostriker:1998fa} 
\be
\dv{E}{t} = -\frac{4\pi G^2 \mdm^2\rho}{\vrel} I
\label{eq:df}
\ee
where $G$ is the gravitational constant, $\rho$ is the gas density and $I$ is the Coulomb logarithm factor. Depending on whether $\vrel$ is larger or smaller than the speed of sound $c_s$ in the gas system, $I$ takes the form of
\begin{equation}
\label{eq:dffactor}
    I =  \Theta(c_s - \vrel) I_1 +  \Theta(\vrel-c_s) I_2,
\end{equation}
where $\Theta$ is the Heaviside function, and $I_1$ and $I_2$ are for the subsonic and supersonic case given respectively by
\begin{equation}
\begin{aligned}
& I_1 =\frac{1}{2} \ln \bigg(\frac{c_s + \vrel}{c_s - \vrel}\bigg) - \frac{\vrel}{c_s}, \\
& I_2 =\frac{1}{2} \ln \bigg(1-\frac{c^2_s}{\vrel^2}\bigg)+\ln \bigg(\frac{R_{\mathrm{sys}}}{\rdm} \bigg),
\end{aligned}
\end{equation}
with $R_{\mathrm{sys}}$ characterizing the spatial size of the gas system.

%The velocity dispersion of both the gas and stars in Leo T is observationally determined to be $\sigma_v \sim 7$ km/s\cite{}. This puts the velocity of gas and DM in Leo T to approximately follow the same Maxwell distribution. For two identical Maxwell distributions, the average relative velocity is equal to $\sqrt{2}$ times the average of individual distribution, and we find $\expval{\vrel} = 15.8$ km/s. By comparison, the speed of sound in the \HI gas is $\sim$ 9 km/s, and the DM objects are therefore mildly supersonic.

The energy lost by the compact DM object due to DF is directly transferred to the gas. To compute the heating rate on gas, we assume the energy fraction of compact DM objects in the entire halo is $f_\DM$, and all of the compact objects have an equal mass $\mdm$ and size $\rdm$ for simplicity. Eq.~\ref{eq:df} then leads to the following volume-averaged heating rate
\be
\label{eq:dfQdot}
\dot{Q}= \frac{12\pi G^2 f_\DM \mdm}{\rwnm^3} \int d\vrel dr \, f(\vrel) \frac{ r^2  \rho_{\mathrm{DM}} \rho_{\mathrm{H}}}{\vrel} I,
\ee
where $\rho_\DM$ and $\rho_{\mathrm{H}}$ are the energy density of DM and \HI respectively. Note that the integration needs to be computed in two parts due to the Heaviside functions in Eq.~(\ref{eq:dffactor}), i.e. an integral in the subsonic regime and another in the supersonic regime. If the DM model features extended distributions of $\mdm$ and $\rdm$, the heating rate in Eq.~(\ref{eq:dfQdot}) needs to be weighted by the mass function and size function.

%Requiring $\dot{Q} \leq \dot{C}$ gives limits on $f_\DM$ as a functionf of $\mdm$ and $\rdm$.

\begin{figure*}[htb]
\begin{tabular}{ll}
\includegraphics[width=0.45\textwidth]{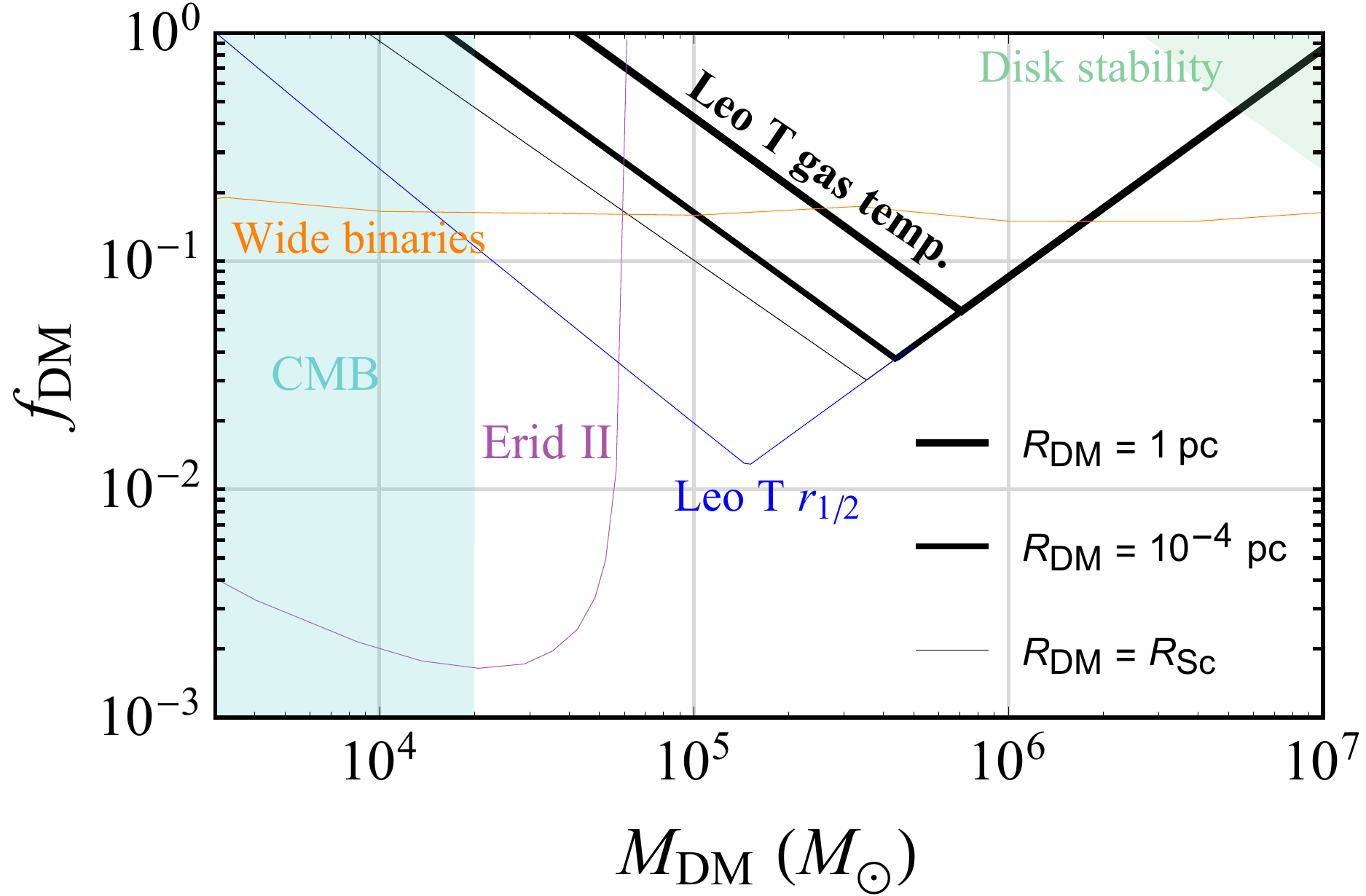}    &  
\includegraphics[width=0.45\textwidth]{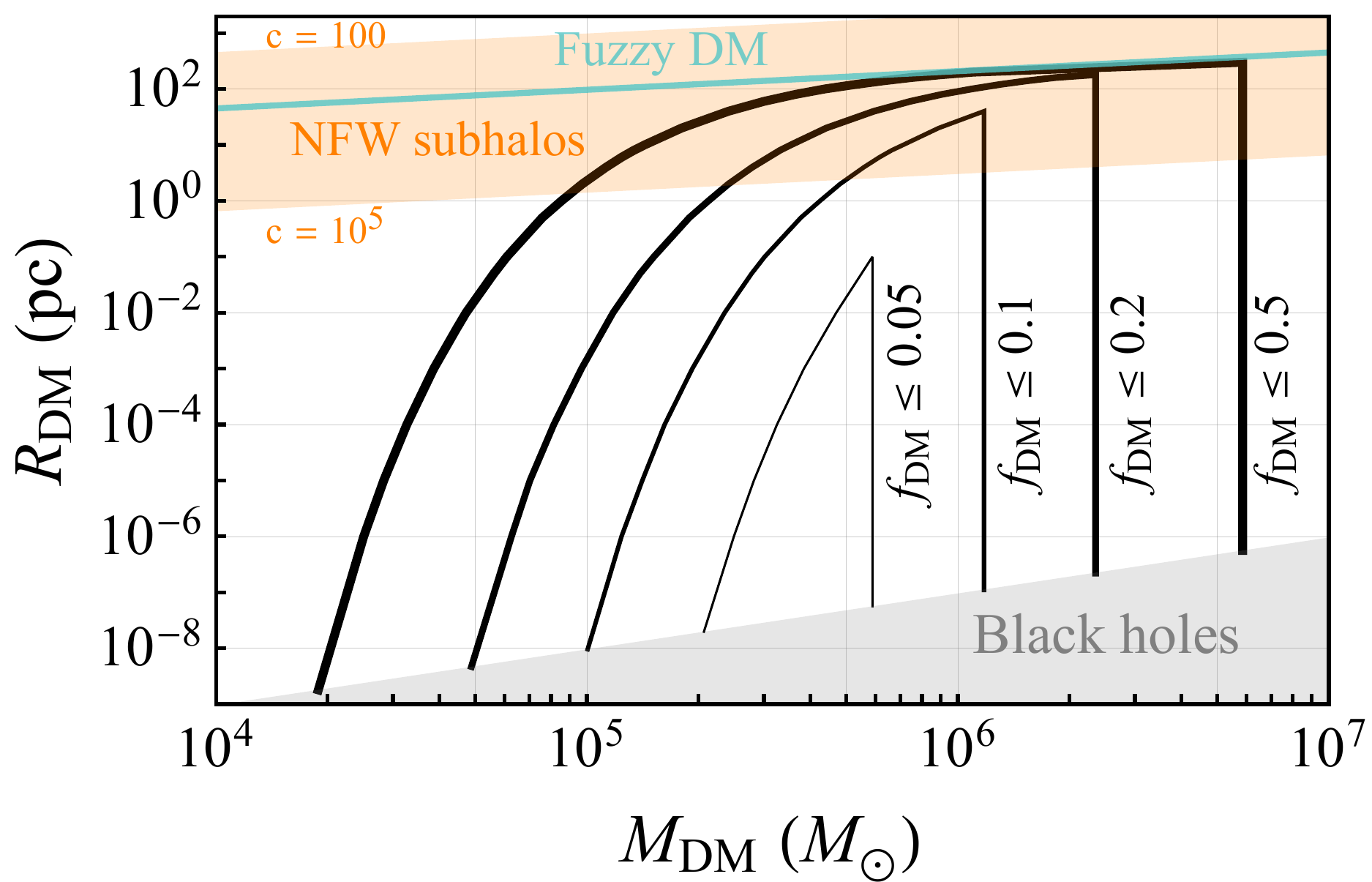}\\
\end{tabular}
    \caption{\textit{Left}: Upper bounds on $f_{\DM}$ as a function of $\mdm$. The black lines are from the gas temperature of Leo T and the thickness of the lines indicate different values of $\rdm$ ranging from the size of black holes to 1 pc. The blue line depicts the limit from the requirement that dynamical heating of stars in Leo T would not increase the half-light radius ($r_{1/2}$) by a factor of two within the stellar halo lifetime. CMB excluded areas~\cite{Ali-Haimoud:2016mbv} are shown in cyan, Erid II limits~\cite{Brandt:2016aco} in violet, the stability of wide binaries~\cite{monroy2014} in orange and galactic disks~\cite{Xu1994} in green. \textit{Right}: The sensitivity of Leo T limits to $\mdm$ and $\rdm$. Black contours are constant values of $f_{\DM}$ probed by Leo T constraints. The orange band depicts the relationship of $\mdm$ and $\rdm$ for NFW subhalos with concentration $10^2 \leq c \leq 10^5$, while the cyan line gives the relationship for granular structures in fuzzy DM scenarios~\cite{Dalal:2022rmp}. }
    \label{fig:df}
\end{figure*}

In the left panel of Fig.~\ref{fig:df}, solid black lines show Leo T gas upper limits on the fraction of compact DM objects. From the thinnest to thickest, we vary $\rdm$ from the Schwarzschild radius to 1 pc. 
%The dashed black line is the Leo T upper limit on PBHs obtained by Refs.~\cite{Lu:2020bmd, TakLu21}, which displays a factor-of-two discrepancy with our limit.
%The discrepancy might arise from the treatment of the weighted integration with respect to $\vrel$. 
At higher $\mdm$, we also impose the ``incredulity'' limit following~\cite{Carr_1999,Lu:2020bmd,TakLu21}, which requires $\mdm \leq f_\DM M_{\mathrm{halo}}$, where $M_{\mathrm{halo}} = 4\pi \int_0^{\rwnm} dr\, r^2 \rho_{\DM}(r)$ is the total DM mass within $\rwnm = 0.35$ kpc. Essentially, this condition ensures there is at least one compact DM object in the environment. In this calculation we have adopted the escape velocity $v_{\mathrm{esc}} = 23.8$ km/s as discussed in Sec.~\ref{sec:property}. Since this estimation of $v_{\mathrm{esc}}$ is largely conservative, we perform another calculation with $v_{\mathrm{esc}} = 62$ km/s to investigate the sensitivity of our results to $v_{\mathrm{esc}}$. We find that the limits are changed by roughly 1\% and thus the uncertainty due to $v_{\mathrm{esc}}$ can be neglected.
The similar constraint from the MW cloud is shown to be above the $f_{\DM} = 1$ baseline~\cite{TakLu21} and therefore we do not show it in Fig.~\ref{fig:df}.

Apart from heating gas in dwarf galaxies, compact DM objects can also heat stellar halos or star clusters via dynamical friction and cause them to expand or dissolve.
Properties of the stellar halo of Leo T (e.g., size, mass, age) have been studied in Refs.~\cite{SimonGeha07, weisz12, Zou21}. We use the methodology of \cite{Brandt:2016aco} and require that the timescale to increase the half-light radius $r_{1/2}$ of Leo T  by a factor of 2 is longer than the lifetime of the stellar halo; this gives us the limit shown in blue in the left panel of Fig.~\ref{fig:df}. The details of the calculations are left to Appendix~\ref{apx:stellar_heating}. 
The reason for stellar limits in Fig.~\ref{fig:df} being stronger than gas limits is that the gas has radiative channels to cool, whereas for the stars, gravitational cooling processes are inefficient \cite{Brandt:2016aco}. This is reflected from the fact that gas cooling lifetimes ($\sim 10^8$ years for Leo T) are typically much shorter than stellar cooling lifetimes (which are typically expected to be longer than Hubble time).

%Otherwise, DF alone would yield stronger limits for higher $\mdm$ because the heating rate Eq.~(\ref{eq:dfQdot}) is proportional to $\mdm$.
%As can be seen in Eq.~(\ref{eq:dfQdot}), the heating rate monotonically increases with $\mdm$, and in principle the limit should always be stronger for heavier DM. The reason for the appearance of a turning point in Leo T limits is that the Leo T halo has a finite total mass. Once the compact DM object is sufficiently heavy, the number of compact objects in the halo become a fractional number smaller than one, which is unphysical. Following~\cite{Lu:2020bmd,Takhistov:2021aqx}, we impose the ``incredulity'' limit $f_\DM \leq \mdm/M_{\mathrm{halo}}$ to require that there is at least one compact object in the halo.

Also shown in Fig.~\ref{fig:df} are the excluded regions that overlap with our Leo T limits, from CMB~\cite{Ali-Haimoud:2016mbv} (cyan), Erid II~\cite{Brandt:2016aco} (violet) [see also \cite{Zou20}], and the stability of wide binaries~\cite{monroy2014} (orange) and galactic disks~\cite{Xu1994} (green). All of these bounds are derived for PBHs only, but can be recast for DM objects with larger $\rdm$. The scaling of CMB limits to $\rdm$ has been investigated by Ref.~\cite{Bai:2020jfm}. The Erid II limit is based on DF and therefore we expect similarly weakened limits for larger $\rdm$ as our Leo T limits. For PBHs in this mass range, there are other limits (see e.g.~\cite{Carr_1999,Inoue:2017csr, LuTak21, Tak21b, Bird:2022wvk}) whose generalization to larger $\rdm$ is currently unexplored.\footnote{Ref.~\cite{Carr_1999} reports strong exclusion limits for PBH masses between $10^6$-$10^{9}\, M_\odot$, requiring $f_{\DM}\leq 10^{-4}$, based on the argument that PBHs would be dragged to galactic nuclei by dynamical friction, increasing the nuclei mass. The calculation depends sensitively to the halo core radius and stellar population~\cite{Carr_1999,Car21,Bird:2022wvk} and therefore we do not show this constraint in Fig.~\ref{fig:df}.}. Projections from future astrometric lensing limits will also provide probes into compact DM in this mass range~\cite{VanTilburg:2018ykj}.

In the right panel of Fig.~\ref{fig:df}, we calculate the contours of constant $f_{\DM}$ constrained by Leo T gas limits on the $\mdm$-$\rdm$ plane. For example, along the $f_{\DM} \leq 0.1$ contour, compact DM objects with these $\mdm$ and $\rdm$ are constrained by Leo T to make up no more than 10\% of the total DM density. The gray region in the bottom indicates the formation of black holes. We also display two exemplary models of DM that feature compact objects and examine if Leo T limits can currently constrain them.
\begin{itemize}
    \item The orange region shows DM subhalos with an NFW profile with concentration $10^2 \leq c \leq 10^5$. In this scenario, $f_{\DM}$ gives the fraction of DM that forms subhalos. CDM subhalos are typically $10\lesssim c \lesssim 100$ as favored by simulations, and are currently not constrained by Leo T at the level of $f_{\DM} \leq 0.5$. Alternate models such as early matter domination~\cite{Blinov:2021axd} and axions~\cite{Fairbairn:2017dmf} can predict subhalos with much higher concentration, and thus be constrained by Leo T limits. Details about the definition of the mass, size and concentration are given in Appendix~\ref{app:nfw}.
    
    \item The cyan line corresponds to $M = 4\pi \rho_0 R^3/3$ where $\rho_0 \simeq 1$ GeV/cm~$^3$ is the average DM density in Leo T. Recent studies of fuzzy DM~\cite{Dalal:2022rmp} with $m_a \sim 10^{-20}$ eV give hints at the formation of granular structures via interference effects. The size of the granule is roughly given by the de Brogile wavelength $R = \order{1} \hbar/(m_a \sigma_v)$ where $\sigma_v$ is the 3-dimensional velocity dispersion of DM ($\simeq 7\sqrt{3}$ km/s for Leo T). The mass of the granule is thus $M = 4\pi \rho_0 R^3/3$. Each granule would behave as a compact object and the abundance of granules $f_{\DM}$ can be potentially constrained by Leo T. For $m_a = 10^{-20}$ eV, the characteristic size of the granule is $27$ pc up to an $\order{1}$ proportionality constant, and the characteristic mass is $5\times 10^3\, M_\odot$ up to the same constant cubed. Currently, these granules are marginally intersecting with the $f_{\DM}\leq 0.2$ contour at $R\sim 100$ pc and $M\sim 10^6$-$10^7\, M_\odot$. Better constraints may be obtained with future study of other gas-rich dwarf galaxies and the improved understanding of granules in fuzzy DM scenarios. It is also worth noting that close to the center of the DM halo (where the \HI gas is the coldest \cite{adams17}), solitons can produce additional strong dynamical heating, but we have not considered this effect. Apart from the granules heating gas, they can also heat the stellar halo of Leo T. Using the methodology given in \cite{Dalal:2022rmp}, observed properties of the stellar halo of Leo T ($\sigma_* = 7$ km/s and $r_{1/2}\sim 170$ pc \cite{Zou21,SimonGeha07}) can be used to add constraints on fuzzy DM in the range $10^{-22}\lesssim m_\mathrm{FDM}\lesssim 10^{-20}$ eV.
\end{itemize}

%%%%%%%%%%%%%%%%%%%%%%%%%%%%%%%%%%%%%%%%%%%%%%%%%%%%%%%%%%%%%
\subsection{Hard sphere scattering}
If non-gravitational interactions between DM and SM exist, more detection strategies become available. To be more specific, we consider heavy DM objects that elastically scatter with SM particles with a cross section set by the geometrical size of DM, $\sigma = \pi \rdm^2$. This is similar to the elastic collision of two hard spheres. When DM passes through a medium of density $\rho$ the energy dissipation rate of DM is~\cite{opik58, DeRujula:1984axn, Bhoonah:2020dzs, Anchordoqui:2021xhu}
\begin{equation}
\label{eq:dEdt}
    \dv{E}{t} = -\rho \sigma \vrel^3.
\end{equation}
%This equation holds when $\mdm$ much greater than nuclei mass. 
This equation is essentially derived based on the scattering rate $n \sigma \vrel$ and the average energy transfer $\sim m\vrel^2$ where $m$ is the particle mass of the medium. Note that the maximum energy transferred is $2m \vrel^2$.

%A very small fraction of the dissipation is converted to visible light~\cite{DeRujula:1984axn}

% The velocity of DM follows a Maxwellian distribution, which we adopt to be~\cite{Lisanti:2016jxe}
% \begin{equation}
% f(\vec{v}) = \begin{cases} \frac{N}{\sigma_v^3} \exp(-\frac{3v^2}{2\sigma_v^2}), \hfill \qif{v < v_{\mathrm{esc}}}, \\ 0, \hfill \qif{v \geq v_{\mathrm{esc}}}, \end{cases}    
% \end{equation}
% where $\sigma_v$ is the velocity dispersion, $v_{\mathrm{esc}}$ the escape velocity, and the normalization constant $N$ chosen such that $\int_0^{v_{\mathrm{esc}}} 4\pi v^2 f(\vec{v}) dv = 1$. In Leo T, $\sigma_v \simeq 7$ km/s and $v_{\mathrm{esc}} \simeq 20$ km/s.

In the interstellar gas mediums considered in this paper, the typical number density of hydrogen is $\sim 0.1/\mathrm{cm}^3$. The recoiled hydrogen particles from DM-hydrogen collision scatters with other hydrogen with a Rutherford cross section $\sim 10^{-17}\, \mathrm{cm}^2$, and therefore the mean free path is $l \sim 1\, \mathrm{pc}$. This is much smaller than the spatial size of the interstellar gas systems that we consider. Therefore, recoiled hydrogen particles can efficiently thermalize with ambient gas particles. Furthermore, because the interstellar gas is sufficiently dilute, the formation of radiation via shock waves~\cite{DeRujula:1984axn} is highly unlikely in this scenario and we expect that the energy loss Eq.~(\ref{eq:dEdt}) is completely converted to heat.
%In one DM-hydrogen collision event, the hydrogen acquires an $\order{\mathrm{keV}}$ kinetic energy.
%the hydrogen atom recoils at a speed twice of the incoming DM object's. For $v_\DM = 300$ (10) km/s, the kinetic energy of the recoiled hydrogen atom is roughly 1 ($10^{-3}$) keV. 
%It then scatters with other hydrogen in the gas with a Rutherford cross section $\sim 10^{-17}\, \mathrm{cm}^2$, and the typical number density of hydrogen is roughly $\sim 0.1/\mathrm{cm}^3$. The mean free path is thus $l \sim 1\, \mathrm{pc}$, and is much less than the spatial size of the gas system $\order{ 100\, \mathrm{pc}}$. Therefore, we expect the recoiled hydrogen particles to thermalize with ambient gas particles and the heat injection is efficient.

To compute the heating rate on gas due to hard sphere scattering, we assume that these compact DM objects compose 100\% of the DM energy density and they have identical mass $\mdm$ and radius $\rdm$. For Leo T, the volume-averaged heating rate is then
\begin{equation}
    \dot{Q} = \frac{\sigma }{\mdm\rwnm^3/3} \int  d\vrel dr\, \vrel^3 f(\vrel) \rho_{\DM}(r) \rho_{\mathrm{H}}(r)  .
\end{equation}
Similarly, we can calculate the heating rate for the MW cloud. As specified in Sec.~\ref{sec:property}, the \HI gas density is taken to be a constant $0.4$/cm${}^{3}$. We adopt the NFW profile for the Milky Way DM halo with $\rho_\chi = \rho_0/[(r/r_0)(1+r/r_0)^2]$ with $\rho_0=0.32$ GeV/cm$^3$, scale radius $r_0=16$ kpc and virial radius $180$ kpc  \cite{Galpy}. This gives DM density $\rho_\DM = 0.64$ GeV/cm$^{3}$ at the location of the cloud ($r= \sqrt{4.68^2 + 1^2}$ kpc).
%and the spatial extent is 4.68 kpc as .
As the heating rate is proportional to $\vrel^3$, the MW cloud turns out to set better limits than Leo T due to higher DM velocity dispersion, although Leo T has a smaller cooling rate.

\begin{figure}[htb]
\includegraphics[width=0.46\textwidth]{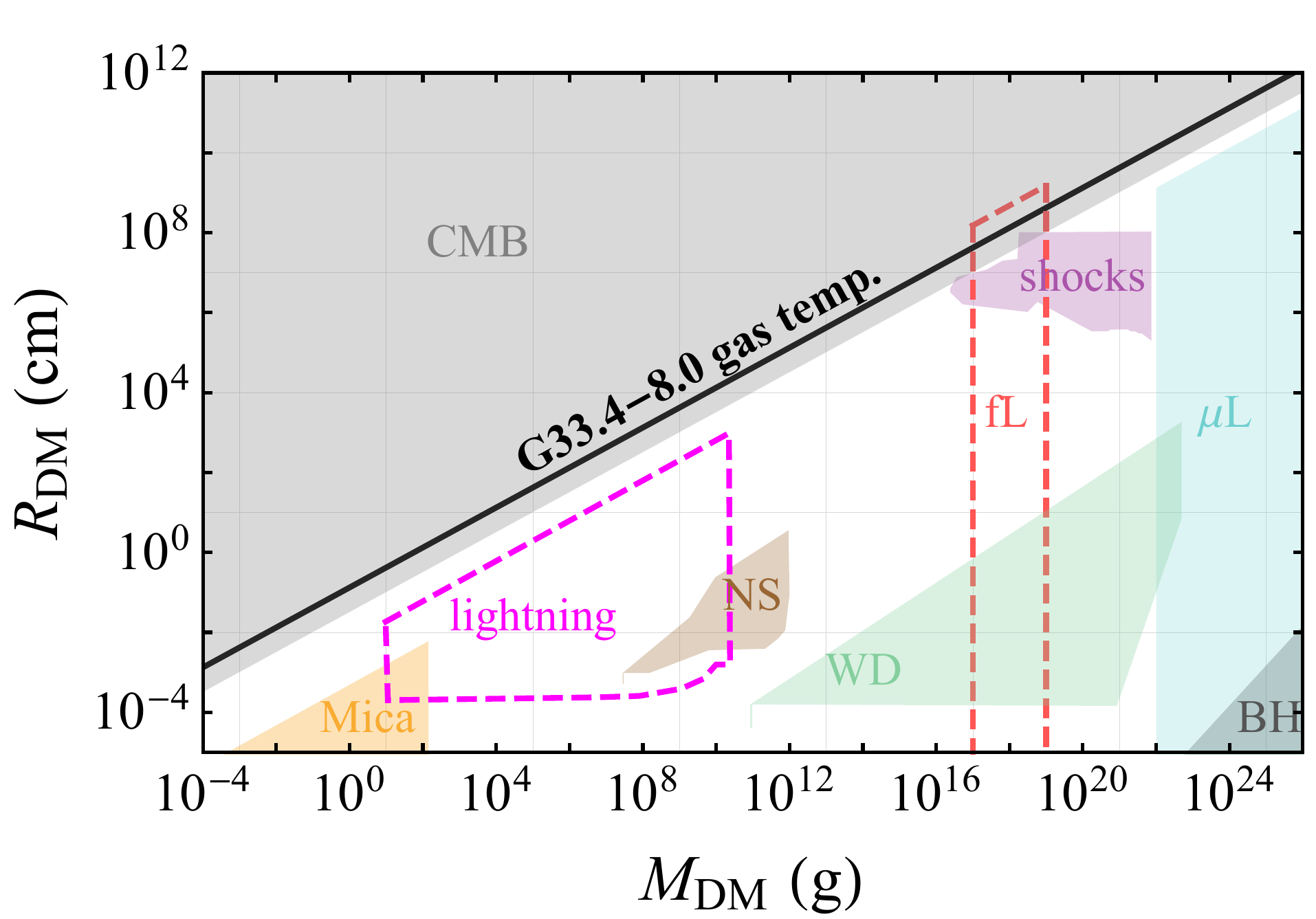}    
    \caption{A compilation of constraints on compact DM objects on radius-mass plane. The black line is the upper limit from the gas temperature of a particular Milky Way cloud G33.4$-$8.0. See the text for other bounds. }
    %The dashed line is recast from the gas heating limit on DM-baryon contact interaction~\cite{Bhoonah:2020dzs}, \new{(see however \cite{Farrar:2019qrv}}). \DW{I think the cloud that Bhoonah et al used is not appropriate so perhaps we can remove their limit from the plot and write about it in the text.}
    \label{fig:geo}
\end{figure}

Our results are shown in Fig.~\ref{fig:geo}. The black line depicts the upper limits on $\rdm$ from the MW cloud. A variety of other limits on $\rdm$ are also included. In the lower right corner, the cyan region is ruled out by microlensing ($\mu L$) observation towards M31~\cite{Niikura:2017zjd,Croon:2020ouk} and the black region denotes the formation of black holes. The dashed pink region can be potentially probed by femtolensing (fL) of gamma-ray bursts~\cite{Jacobs:2014yca,barnacka12,Katz:2018zrn}, but the validity is subject to further investigation of finite-source effects~\cite{Katz:2018zrn}. These bounds are purely gravitational and do not assume a DM-baryon interaction. Other limits are derived from various constraints on DM-baryon scattering cross sections, including CMB (gray)~\cite{Dvorkin:2013cea,Jacobs:2014yca}, Mica (orange)~\cite{Price:1986ky,Jacobs:2014yca}, neutron stars (brown) and white dwarfs (green)~\cite{Graham:2018efk, SinghSidhu:2019tbr, Dessert:2021wjx}, observability of shock waves from DM-star collisions (violet)~\cite{Das:2021drz}, and lightning (dashed magenta)~\cite{Sidhu:2019fgg,Starkman:2020sbz} (see however~\cite{Cooray:2021dvp}). We also refer readers to~\cite{Dhakal:2022rwn} for a new study in using meteor radars to constrain DM-nuclei scattering.
% Two additional constraints shown in dashed should be considered with caution. In the dashed magenta area, lightning events could be observed when DM traverses the atmosphere~\cite{Sidhu:2019fgg,Starkman:2020sbz} (see however~\cite{Cooray:2021dvp}). DM in the dashed red region is sensitive to femtolensing towards gamma-ray burst sources~\cite{Jacobs:2014yca,barnacka12,Katz:2018zrn}, but the validity of this probe is subject to further investigation of finite-source effects~\cite{Katz:2018zrn}.

We note that Ref.~\cite{Bhoonah:2020dzs} reports a similar bound based on another MW-environment gas cloud, G357.8-4.7-55. This gas cloud has a lower cooling rate $3.4\times 10^{-28}$ $\mathrm{erg}\, \mathrm{cm}^{-3}\, \mathrm{s}^{-1}$ and a larger DM density $17$ GeV/cm$^{3}$ (cf. the cooling rate and DM density for G33.4-8.0 are $2.1\times 10^{-27}$ $\mathrm{erg}\, \mathrm{cm}^{-3}\, \mathrm{s}^{-1}$ and $0.64$ GeV/cm$^{3}$). In consequence, their limits purport to be stronger than CMB. However, as discussed in Refs.~\cite{Farrar:2019qrv,WadFar21}, G357.8-4.7-55 is immersed in an extreme environment (i.e., in the hot and high-velocity outflow from the Galactic Center, with $T_\mathrm{outflow}\sim10^{6-7}$ K), raising questions on if this gas cloud is in a steady state in order to derive constraints on DM heating. 
We show the close-up of the comparison between these two limits as well as limits from Leo T and self-interacting DM in Appendix~\ref{app:closeup}.

% In the calculation above, we have neglected the decrease in DM velocity due to the dissipation Eq.~(\ref{eq:dEdt}). Along the path of a compact DM object, the change in its kinetic energy is given by $\Delta E = \int dx\, dE/dx$ where $dE/dx = (dE/dt)/\vrel$ is the stopping power. The total loss of kinetic energy is thus $\Delta E \sim R_{\mathrm{system}} \times dE/dx$. We can therefore estimate the ratio
% \begin{equation}
% \begin{aligned}
%     \frac{\Delta E}{\mdm \vrel^2/2} \sim & 3.4\times10^{-2}\qty(\frac{\rho}{\mathrm{GeV/cm}^3})\qty(\frac{R_{\mathrm{system}}}{\mathrm{kpc}}) \\
%     &\qty(\frac{\rdm}{\mathrm{cm}})^2 \qty(\frac{\mathrm{g}}{\mdm}).
% \end{aligned}
% \label{eq:deltav}
% \end{equation}
% (Refine the comment below.)
%As long as the ratio is small enough (which sets a condition on $\rdm^2/\mdm$), we can safely neglect the effect on the change in the velocity distribution of DM. On the other hand, if this ratio is $\order{1}$, DM would be confined in the gas system and and this region on the $\rdm$-$\mdm$ space is ruled out by the DM profile. 

%%%
% \subsection{Axion quark nugget}

% The emission rate of photons with energy $E_\gamma$ in unit volume is
% \begin{equation}
%     \dv{N(E_\gamma)}{tdV} = \kappa(E_\gamma) \rho_\DM(r) \rho(r) \expval{\vrel},
% \end{equation}
% where $\kappa(E_\gamma)$ is proportional to the probability that the annihilation produces a photon with energy $E_\gamma$ and depends only on the property of AQN.

\section{Discussion \& Conclusions}

Observations of cold and metal-poor interstellar gas systems can be a great complement to the program of DM direct and indirect detection. Requiring the heat injection rate from DM lower than the astrophysical cooling rate of the gas can yield compelling limits on a variety of DM models. In this paper, we have derived limits on the following scenarios:
\\$(i)$ we place upper limits on the electron coupling of axion DM for $m_a < 100$ keV. This constraint evades the overburden effect that laboratory direct detection experiments suffer from, and rules out the space of large couplings.
\\$(ii)$ we constrain the abundance of compact DM objects in the mass range $10^4 \mathrm{-} 10^7\, M_{\odot}$. We also show the sensitivity to the spatial extent of the compact object. This limit is purely derived from dynamical friction between the compact object and gas and is thus robust for any type of compact objects.
\\$(iii)$ for DM-nuclei scattering that saturates the geometric cross section, we find upper bounds on the radius of the composite DM state.
\\$(iv)$ finally, we set upper limits on the abundance of DM in the form magnetically charged black holes.

% End with a note on future probes of gas-rich dwarf galaxies and simulation of DM-gas interaction?

For calculating the DM bounds from Leo T, we used gas and DM profiles from the model of Leo T by Ref.~\cite{faerman13}. Note however that their model assumes the gas is in hydrostatic equilibrium (i.e., the gravitational force due to the DM halo is balanced by the gas thermal pressure). Their model also does not take into account astrophysical heating and radiative cooling of the \HI gas.
In a future study, we plan to perform hydrodynamic simulations of gas-rich dwarfs like Leo T which include thermal feedback from non-standard DM alongside the standard astrophysical heating and cooling effects. It will also be interesting to perform simulations of Milky Way \HI clouds including DM heating.

Let us now discuss observational prospects of ultra-faint \HII-rich dwarf galaxies similar to Leo T.
Numerous ongoing and upcoming surveys will be able to find and characterize such dwarfs (e.g., 21cm surveys like WALLABY \cite{Kor20_Wallaby}, MeerKAT \cite{Mad21}, Apertif \cite{van22}, FAST \cite{Zha21}, SKA \cite{SKA}, and optical surveys like DESI \cite{DESI}, HSC \cite{Aih18}, Dragonfly \cite{Dan20}, Rubin observatory \cite{LSST,Drl19_LSST,Mut21}, Roman telescope \cite{WFIRST}). Rubin observatory will likely be the most impactful in this regard due to its wide field of view, and its sensitivity to detect dwarfs with brightness similar to Leo T (i.e., $M_V=-8$) up to $\sim5$ Mpc \cite{Mut21}. This opens a possibility of detecting hundreds of galaxies similar to Leo T and could enable more stringent probes of heat exchange due to DM.

%%%
%\section{Charged Planck-scale relics/CHAMPs}

%\cite{Lehmann:2019zgt}

%%%
%\subsection{Electromagnetic injections from ECO binary mergers}

%\cite{Hertzberg:2020dbk,Amin:2020vja,Kritos:2021nsf}

%%%
%\section{CHAMPs}

\begin{acknowledgments}
We thank Mellisa Diamond, Glennys Farrar, Chris Hamilton, Ken Van Tilburg, Scott Tremaine, Huangyu Xiao and Tomer Yavetz for useful discussions.
We also thank Yakov Faerman and Shmuel Bialy for providing us the data for the best-fit model of Leo~T from Ref.~\cite{faerman13}.
DW gratefully acknowledges support from the Friends of the Institute for Advanced Study Membership and from the W. M. Keck Foundation Fund.
\end{acknowledgments}

\appendix

\section{Cooling rates}
\label{app:cooling}
We had shown the cooling rates of different ISM phases of the Milky Way in Fig.~\ref{fig:coolrate}. In this Appendix, we discuss the properties of ISM phases used in our cooling rate calculations. We calculate the cooling rate using Eq.~\ref{eq:grackle} and used the properties in Table~\ref{tab:coolrate} as input. For the metallicity, we use [Fe/H]$\sim 0$ for all Milky Way systems. It is worth mentioning that the molecular cloud (MC) parameters that we show are for diffuse H$_{2}$ systems (the radiative cooling rate is much larger for dense H$_{2}$ systems).
%\vspace{-0.2in}

%\section{DM velocity distribution}

\section{Axioelectric heating rate}
\label{app:axion}

\begin{table}
\begin{ruledtabular}
\begin{tabular}{llllll}
Astrophysical & n & T  & Cooling rate \\ 
medium & (cm$^{-3}$) & (K) & (erg cm$^{-3}$ s$^{-1}$)\\ \hline
WNM (Leo T)  & 0.06 & 6100 & $7 \times 10^{-30}$\\
CNM (G33.4$-$8.0) & 0.4 & 400 & $1.46 \times 10^{-27}$\\
MC (MW)  & 100 & 50 & $3.16 \times 10^{-24}$\\
CNM (MW)  & 30 & 100 & $2.02 \times 10^{-24}$\\
WNM (MW)  & 0.6 & 5000 & $8.3 \times 10^{-27}$\\
WIM (MW)  & 0.3 & $10^4$ & $5.4 \times 10^{-26}$\\
HIM (MW)  & 0.003 & 10$^6$ & $1.3 \times 10^{-27}$\\
% G16.0$+$3.0       & 1.7$\pm$0.2 & 480$\pm$20 & 2.34$\pm$0.2 & 0.43$\pm$0.05 \\
% G26.9$-$6.3       & 2.5$\pm$0.5 & 200$\pm$13 & 3.85$\pm$0.36 & 0.84$\pm$0.19 \\
\end{tabular}
\end{ruledtabular}
\caption{\label{table:clouds}Properties of systems used in calculating the cooling rates shown in Fig.~\ref{fig:coolrate}. These values are input in Eq.~\ref{eq:grackle} to calculate the cooling rate.
The data for different media in the Milky Way (MW) are taken from Table 1.3 of Ref.~\cite{Dra11}. The MW phases are labelled as MC (molecular clouds), CNM (cold neutral medium), WNM (warm neutral medium), WIM (warm ionized medium), HIM (hot ionized medium). Note that we only show typical averaged values of the properties of MW phases; these numbers can be different for different systems in a particular phase (e.g., different clouds in WNM of MW can have different properties, e.g., see \cite{Leh04}).}
\label{tab:coolrate}
\end{table}

\begin{figure}
\centering
\includegraphics[scale=0.6,keepaspectratio=true]{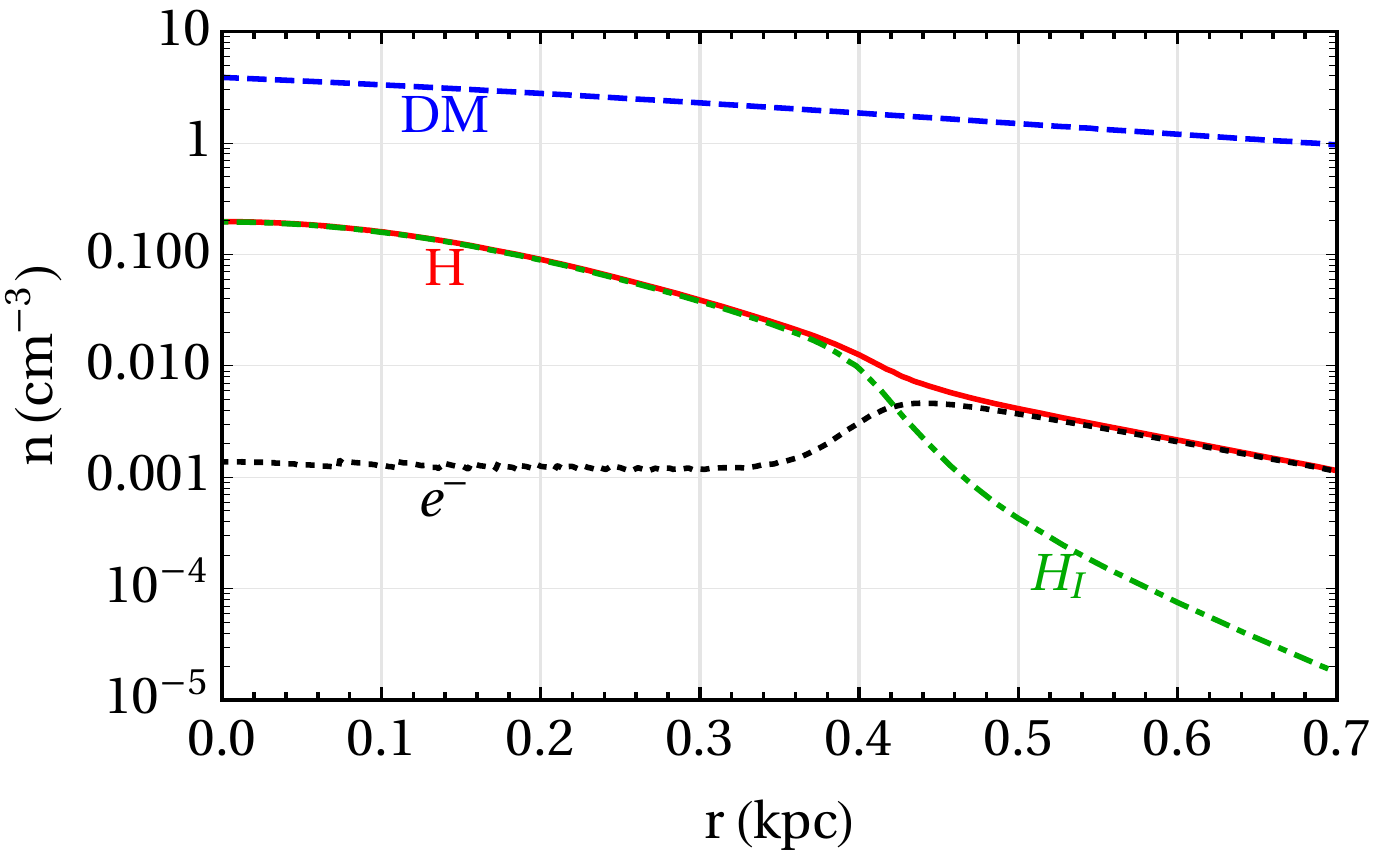}
\caption{We use a model of the gas-rich Leo~T dwarf galaxy by Ref.~\cite{faerman13}, which was fitted to 21cm measurements of the galaxy by \cite{leoObsOld08}, and is also consistent with recent stellar velocity dispersion estimates by \cite{Zou21}.
We show the number density of DM (for $m_\mathrm{DM}=1$ GeV), atomic hydrogen (H\textsc{i}), electrons (e$^-$) and total hydrogen (H) components of the model.
This figure is taken from Ref.~\cite{WadFar21} and is shown here for self-contained discussion.} 
\label{fig:LeoTobs}
\end{figure}

In this appendix we give more details of Eq.~(\ref{eq:axionheat}). The radial distribution of DM density and $\nH$ in Leo T are determined by Ref.~\cite{faerman13} and are plotted by Fig.~1 in Ref.~\cite{WadFar21}. For self-contained discussion, we show the Leo T density profile in Fig.~\ref{fig:LeoTobs}. The heating efficiency function $f_e$ for electrons with kinetic energy $\omega$ takes the form~\cite{Kim20}

\be\begin{split}
f_e(\omega)\simeq\, & 1-(1-x_e^{0.27})^{1.32}\\
& +3.98 \bigg(\frac{11\mathrm{eV}}{\omega}\bigg)^{0.7}x_e^{0.4} (1-x_e^{0.34})^2,
\label{eq:fe}\end{split}\ee
where $x_e = n_e/\nH$ is the ionization fraction and in Leo T, $x_e \simeq 0.02$.

\begin{figure}[htb]
    \centering
    \includegraphics[width=0.45\textwidth]{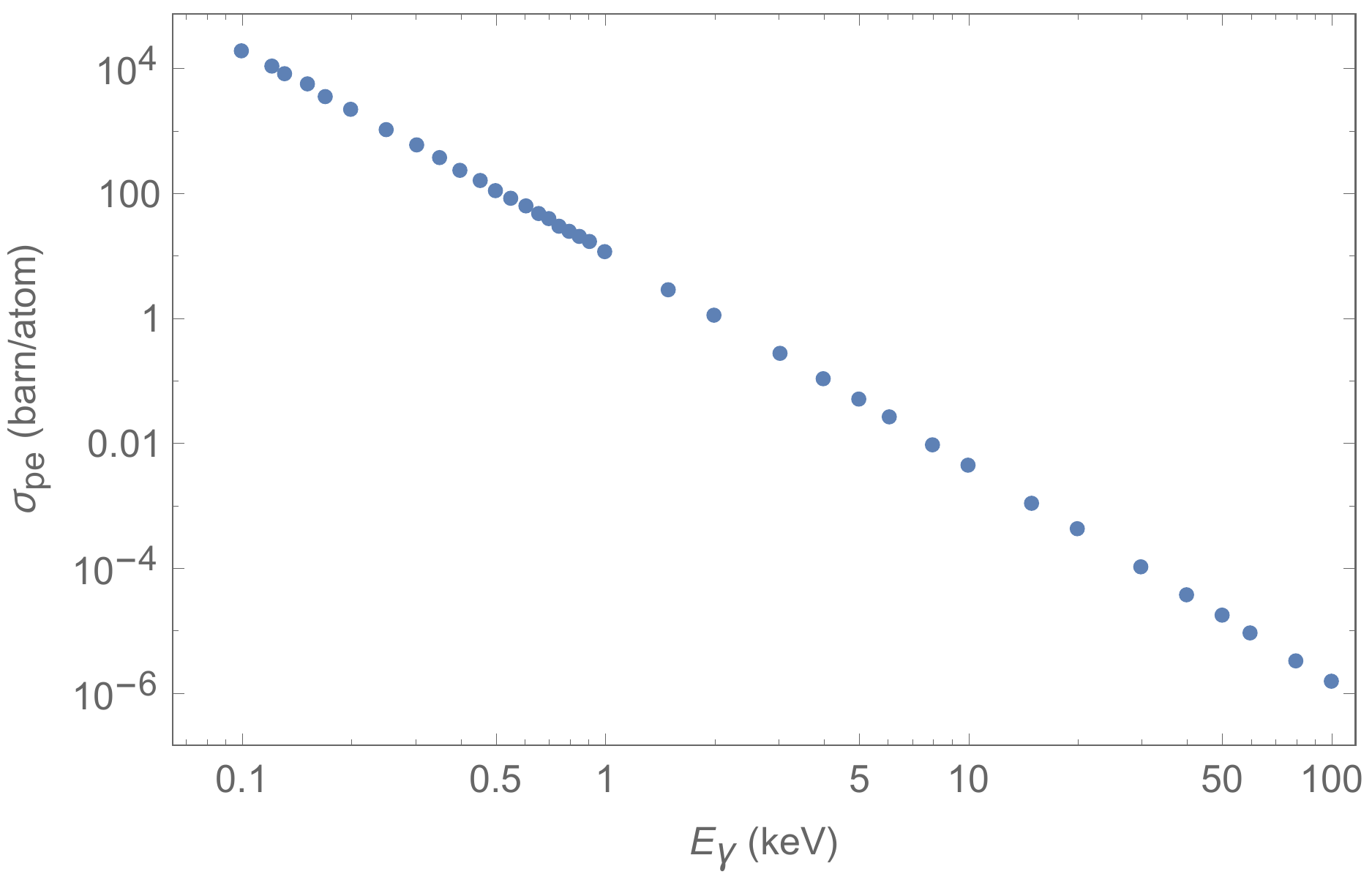}
    \caption{Photoelectric cross section $\sigma_{pe}$ for hydrogen. Data taken from Ref.~\cite{VEIGELE197351}.}
    \label{fig:sigmape}
\end{figure}

In the non-relativistic limit, the axioelectric cross section is~\cite{dimopoulos86,pospelov08}
\begin{equation}
    \sigma_{ae} = \sigma_{pe} \frac{g_{ae}^2}{\vrel} \frac{3m_a^2}{16\pi \alpha m_e^2},
\end{equation}
where $\sigma_{pe}$ is the the photoelectric cross section of the same atom and $\alpha=1/137$ is the electromagnetic fine structure constant. We obtain $\sigma_{pe}$ of hydrogen from Ref.~\cite{VEIGELE197351} and plot it in Fig.~\ref{fig:sigmape}.
Note that the inverse proportionality of $\sigma_{pe}$ with $\vrel$ makes the heating rate Eq.~(\ref{eq:axionheat}) independent of $\vrel$. Because of this, Leo T gives stronger limits than the MW cloud.

%Finally, we point out that the heating rate Eq.~(\ref{eq:axionheat}) should be additionally suppressed by the opacity factor of the gas~\cite{Wadekar:2021qae}.

\section{NFW subhalos}
\label{app:nfw}
We parametrize the density profile of NFW subhalos with scale radius $r_s$ and concentration $c$~\cite{Ramani:2020hdo}
\begin{equation}
    \rho(r) = \frac{200c^3 \rho_c}{3(\ln(1+c)-c/(1+c))} \frac{1}{r/r_s(1+r/r_s)^2},
\end{equation}
where $\rho_c$ is the critical density of the Universe. As the spatial integration of the density is formally divergent, we cut off the profile at $R = 100\, r_s$ and obtain the mass of the subhalo by $M = 4\pi \int_0^{100\, r_s} dr\, r^2\rho(r) $. Based on these we can thus establish the relationship between $M$ and $R$ for fixed values of $c$. We note that in studies of boson stars, the spatial size is often taken to be $R_{90}$ which encloses 90\% of the mass. For NFW profile, $R_{90}$ is roughly at $69 r_s$ and would raise an $\order{1}$ correction to the definition of the size.

\section{G357.8-4.7-55 limits on hard sphere scattering}
\label{app:closeup}

\begin{figure}
    \centering
    \includegraphics[width=0.45\textwidth]{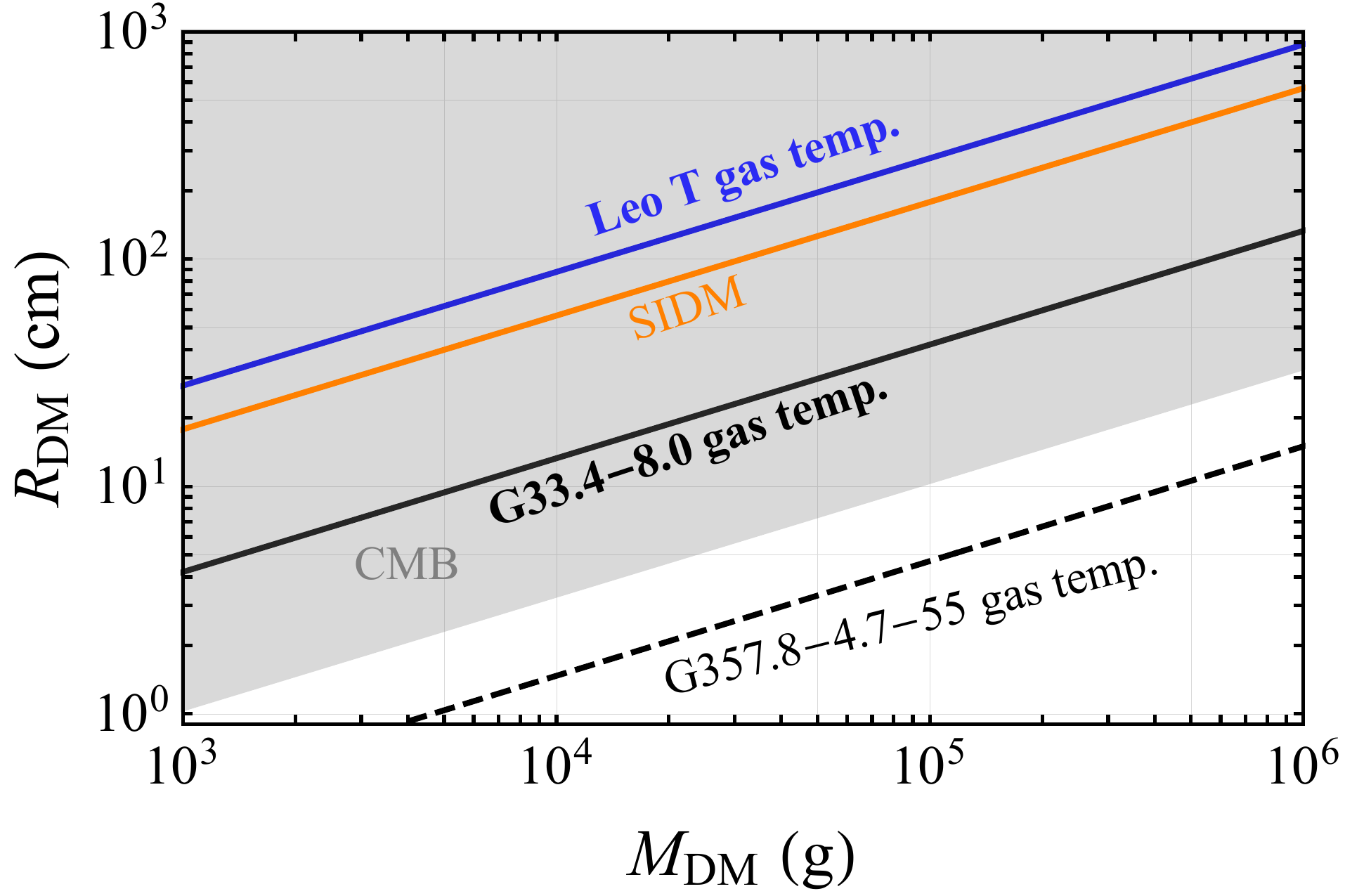}
    \caption{A close-up of Fig.~\ref{fig:geo} for $\mdm$ between $10^3$ and $10^6$ g. The orange line corresponds to the upper bound on self-interacting DM cross section $\lesssim 1\, \mathrm{cm}^2/\mathrm{g}$~\cite{Tulin:2017ara}.}
    \label{fig:closeup}
\end{figure}

Fig.~\ref{fig:closeup} shows the close-up of Fig.~\ref{fig:geo} in the $10^3\mathrm{-}10^6$ g mass range. The dashed line is recast from the gas heating limit on DM-baryon contact interaction~\cite{Bhoonah:2020dzs} based on G357.8-4.7-55. As pointed out by Ref.~\cite{Farrar:2019qrv}, this gas cloud is inappropriate for placing limits on DM heat injection. We further display a few additional limits. The blue line shows the limit from Leo T gas temperature with $v_{\mathrm{esc}} = 23.8$ km/s. Changing the escape velocity to 62 km/s improves the Leo T limit by roughly 3\%. If DM states also self-scatter with a geometrical cross section, the self-interaction is well constrained by astrophysical measurements at the level of $\lesssim 1 \, \mathrm{cm}^2$ per DM mass in gram~\cite{Tulin:2017ara}. This translates to the orange line.

\section{Heating of stellar halo in Leo T}
\label{apx:stellar_heating}

In section~\ref{sec:df}, we presented limits from dynamical friction heating of stellar halo of Leo T due to compact DM. Here, we briefly show the steps involved in our calculation of the limits. Note that we have closely followed the methodology of Ref.~\cite{Brandt:2016aco} and encourage the reader to refer to their paper for further details. \cite{Brandt:2016aco} derived their limits from heating of a particular stellar cluster at the center of Eridanus II, whereas here we consider heating of the stellar halo in Leo T. Ref.~\cite{SimonGeha07} fit a Plummer profile to the stellar halo of Leo T and inferred $r_{1/2} = 170\pm 15$ pc. Ref.~\cite{weisz12} estimated that the half of the stellar mass in Leo T formed prior to $7.6$ Gyr, so we use that period as the stellar halo lifetime ($t_{1/2}$) in our calculations.

Due to dynamical friction by compact DM objects, the half-light radius ($r_{1/2}$) of the stellar halo increases at a rate \cite{Brandt:2016aco}
\begin{equation}
\frac{dr_{1/2}}{dt} = \frac{4\sqrt{2}\pi G f_\mathrm{DM} M_\mathrm{DM}}{2\beta \sigma_\mathrm{DM} r_{1/2}} \ln \Lambda
\end{equation}
where $\sigma_\mathrm{DM}$ ($M_\mathrm{DM}$) is the 3D velocity dispersion (mass) of compact DM objects. $f_\mathrm{DM}$ is the fractional contribution of compact objects to the total DM mass density.
The Coulomb logarithm is given by \cite{Brandt:2016aco}
\begin{equation}
\ln \Lambda \sim \ln \left( \frac{r_{1/2}\sigma^2}{G M_\mathrm{DM}}\right)
\end{equation}
where we assumed that DM objects are much heavier than stars.
We conservatively use $\beta \sim 10$ estimated for a cored Sersic profile \cite{Brandt:2016aco}.
 We require that $r_{1/2}$ to not increase by more than a factor of 2 within the lifetime of Leo T, which gives the following limit on the fraction of DM allowed as compact objects
 
\begin{equation}
\begin{split}
f_\mathrm{DM}\lesssim\, &0.02 \left(\frac{r_{1/2}}{170\, \mathrm{pc}}\right)^2 \left(\frac{t_{1/2}}{7.6\, \mathrm{Gyr}}\right)^{-1} \left(\frac{M_\mathrm{DM}}{10^5\, M_\odot}\right)^{-1}\\ & \times \left(\frac{\sigma_\mathrm{LOS}}{7.6\, \mathrm{km/s}}\right) \left(\frac{\beta}{10}\right) \left(\frac{\ln \Lambda}{9.86}\right)^{-1}
\end{split}
\end{equation}
where $\sigma_\mathrm{LOS}$ is the 1D velocity dispersion (along line of sight).

\bibliography{apssamp}% Produces the bibliography via BibTeX.

\end{document}